\documentclass[12pt,a4paper]{article}

\usepackage[latin1]{inputenc}
\usepackage{amsmath}
\usepackage{amsfonts}
\usepackage{graphicx}
\usepackage{verbatim}
\usepackage[dvips]{color}
\usepackage{multirow}
 
\renewcommand{\baselinestretch}{1.3}

\makeatletter
\@addtoreset{equation}{section}

\makeatother

\newcounter{Fig}[figure]

\newcounter{Tab}[table]

  {\refstepcounter{Tab}
   \addtocounter{table}{1}
   \samepage\vspace{0.2cm}
   \centerline{#1} \nobreak
   \begin{verse}{Table \thesection.\arabic{Tab}:}
  }%
  {\end{verse} \vspace{0.2cm}}

\relax

\newcommand{\bqa}{\begin{eqnarray*}}
\newcommand{\eqa}{\end{eqnarray*}}
\newcommand{\bqan}{\begin{eqnarray}}
\newcommand{\eqan}{\end{eqnarray}}
\newcommand{\bqt}{\begin{quote}}
\newcommand{\eqt}{\end{quote}}
\newcommand{\bt}{\begin{tabbing}}
\newcommand{\et}{\end{tabbing}}
\newcommand{\bit}{\begin{itemize}}
\newcommand{\eit}{\end{itemize}}
\newcommand{\ben}{\begin{enumerate}}
\newcommand{\een}{\end{enumerate}}
\newcommand{\beq}{\begin{equation}}
\newcommand{\eeq}{\end{equation}}
\newcommand{\bdefi}{\begin{definition}}
\newcommand{\edefi}{\end{definition}}
\newcommand{\bpro}{\begin{proposition}}
\newcommand{\epro}{\end{proposition}}
\newcommand{\blem}{\begin{lemma}}
\newcommand{\elem}{\end{lemma}}
\newcommand{\bth}{\begin{theorem}}
\newcommand{\eth}{\end{theorem}}
\newcommand{\bco}{\begin{corollary}}
\newcommand{\eco}{\end{corollary}}
\newcommand{\bdes}{\begin{description}}
\newcommand{\edes}{\end{description}}
\newcommand{\bre}{\begin{remark}}
\newcommand{\ere}{\end{remark}}

\newtheorem{definition}{Definition}[section]
\newtheorem{proposition}[definition]{Proposition}
\newtheorem{lemma}[definition]{Lemma}
\newtheorem{theorem}[definition]{Theorem}
\newtheorem{corollary}[definition]{Corollary}
\newtheorem{remark}[definition]{Remark}

\def\bib{\par\noindent\hangindent=0.5 true cm\hangafter=1}

\newcommand{\m}{\hspace{-0.25 cm}}
\newcommand{\si}{\sigma}
\newcommand{\ep}{\varepsilon}

\newcommand{\sumi}{\sum_{i=1}^n}
\newcommand{\sumj}{\sum_{j=1}^n}

\newcommand{\Var}{\mbox{Var}}
\newcommand{\Cov}{\mbox{Cov}}

\newcommand{\ds}{\displaystyle}

\newcommand{\eps}{\varepsilon}
\newcommand{\Y}{\mathcal{Y}}
\newcommand{\F}{\mathcal{F}}
\newcommand{\G}{\mathcal{G}}

\begin{document}

\begin{titlepage}
\title{\bf Heteroscedastic semiparametric transformation models: estimation and testing for validity}
\author{{\large Natalie N\textsc{{eumeyer}}\footnote{Department of Mathematics, University of Hamburg, Bundesstrasse 55, 20146 Hamburg, Germany, E-mail: neumeyer@math.uni-hamburg.de}}
\and
{\large Hohsuk N\textsc{{oh}}\footnote{Department of Statistics, Sookmyung Women's University, 100 Cheongpa-ro 47-gil, Yongsan-gu,
Seoul, South Korea 140-742, E-mail: word5810@gmail.com}} 
\and
 {\large Ingrid V\textsc{{an} K{eilegom}}\footnote{Institute of Statistics, Universit\'e catholique de Louvain, Voie du Roman Pays 20, 1348 Louvain-la-Neuve, Belgium, E-mail: ingrid.vankeilegom@uclouvain.be}} \\[.5cm]}

\maketitle

\renewcommand{\baselinestretch}{1.1}

\begin{abstract}
In this paper we consider a heteroscedastic transformation model of the form $\Lambda_{\vartheta}(Y) = m(X) + \sigma(X) \eps$, where $\Lambda_{\vartheta}$ belongs to a parametric family of monotone transformations, $m(\cdot)$ and $\sigma(\cdot)$ are unknown but smooth functions, $\eps$ is independent of the $d$-dimensional vector of covariates $X$, $E(\eps)=0$ and $\mbox{Var}(\eps)=0$.  In this model, we first consider the estimation of the unknown components of the model, namely $\vartheta$, $m(\cdot)$, $\sigma(\cdot)$ and the distribution of $\eps$, and we show the asymptotic normality of the proposed estimators.  Second, we propose tests for the validity of the model, and establish the limiting distribution of the test statistics  under the null  hypothesis.  A bootstrap procedure is proposed to approximate the critical values of the tests.  Finally, we carry out a simulation study to verify the small sample behavior of the proposed estimators and tests.
\end{abstract}

\vspace*{.5cm}

\noindent{\bf Key words:} bootstrap, empirical distribution function, empirical independence process, local polynomial estimator, location-scale model, model specification, nonparametric regression, profile likelihood estimator 

\vspace*{.5cm}

\noindent{\bf AMS 2010 classification:} 62G05, 62G08, 62G09, 62G10, 62G30

\end{titlepage}

\small
\normalsize
\addtocounter{page}{1}

\section{Introduction}  \label{introduction}

Assume we observe independent copies of a random vector $(X,Y)$, where $X$ represents a $d$-dimensional covariate and $Y$ is a univariate response. One possibility is to analyze these data by fitting a non- or semiparametric regression model, i.\,e.\
\begin{equation}\label{mod1}
Y=m(X)+\varepsilon, \mbox{ where } E[\varepsilon\mid X]=0.
\end{equation}
Doing so, often the conditional error distribution, given the covariate, still depends on $X$, which means that the dependency of the response $Y$ on the covariate $X$ goes beyond the first moment. If only the second moment is dependent on $X$ one can fit a nonparametric location-scale model of the form 
\begin{equation}\label{mod2}
Y=m(X)+\sigma(X)\varepsilon, \mbox{ where } \varepsilon \perp X \mbox{ with } E[\varepsilon]=0, \mbox{Var}(\varepsilon)=1.
\end{equation}
Here and throughout the paper $Z\perp X$ means that $Z$ and $X$ are stochastically independent. 
Such nonparametric location-scale models have been widely used, see e.\,g.\ Akritas and Van Keilegom (2001), Dette, von Lieres und Wilkau and Sperlich (2005) or Hu\v{s}kov\'{a} and Meintanis  (2010), among many others.  
Note that the conditional normal distribution is always a special case because from $Y|X=x \sim N(m(x),\sigma^2(x))$ it follows that  $\eps\sim N(0,1)$ does not depend on $X$. 
The general location-scale model (\ref{mod2}) has several advantages over the unstructured model (\ref{mod1}). First, the asymptotic analysis of statistical procedures often simplifies a lot. Further, the model allows to estimate the error distribution with a parametric $\sqrt{n}$-rate, see Akritas and Van Keilegom (2001). Therefore the estimation of the conditional distribution of $Y$ given $X$ is much more efficient.
Goodness-of-fit as well as other specification tests have been developed that specifically use the location-scale structure, see Section 2.4 in the recent review by Gonz\'{a}lez-Manteiga and Crujeiras (2013). 
When data $(X,Y_1,Y_2)$ have been observed and one's interest lies in the dependence between $Y_1$ and $Y_2$, given $X$, under the location-scale structure the conditional copula of $(Y_1,Y_2)$, given $X$, can not only  be estimated with  $\sqrt{n}$-rate, but also as precisely as if the errors would be known, see Gijbels, Omelka and Veraverbeke (2013). 
  
The construction of valid resampling procedures is essential for most hypothesis tests in nonparametric regression.  It is known that in heteroscedastic regression models simple residual bootstrap methods generally do not lead to valid procedures. Thus mostly wild bootstrap is used, see H\"{a}rdle and Mammen (1993) and Stute, Gonz\'{a}lez Manteiga and Presedo Quindimil (1998).   
However,  Zhu, Fujikoshi and Naito (2001) show that wild bootstrap may fail if the conditional 4th moment of the error distribution depends on the covariate, while for the procedure considered there it works in the location-scale context. There are other cases where wild bootstrap even fails in the location-scale  model (\ref{mod2}), see e.\,g.\ Neumeyer and Sperlich (2006). A (smooth or not smooth) heteroscedastic residual bootstrap often can be an alternative, see Neumeyer (2009a), and explicitly makes use of the location-scale structure.

Before application of model (\ref{mod2}) a specification test should be conducted, i.\,e.\ a test for independence of $\varepsilon$ and $X$. Such tests have been suggested by Einmahl and Van Keilegom (2008), Neumeyer (2009b), and Hl\'{a}vka, Hu\v{s}kov\'{a} and Meintanis (2011).
However, if those tests reject the null hypothesis a remedy might be to transform the response $Y$ by a suitable transformation $\Lambda$ before fitting the location-scale model to the data $(X,Y)$. 

It is very common in practice to transform the response variable before fitting a regression model to the data. The aim of the transformation is to reduce skewness or heteroscedasticity, or to induce normality. 
Often the transformation is chosen from a parametric class such as the famous class of Box-Cox power transformations introduced by 
Box and Cox (1964). Generalizations of this class were suggested by Bickel and Doksum (1981) and Yeo and Johnson (2000), among others. 
The parameter of the transformation in the class can be chosen data dependently by a profile likelihood approach, for instance. 
There is a huge literature on parametric transformation models and we refer to the monograph by Carroll and Ruppert (1988); see also the references in Fan and Fine (2013). 
Nonparametric estimation of the transformation in the context of parametric regression models has been considered by Horowitz (1996) and Zhou, Lin and Johnson (2008),  among others. Horowitz (2009) reviews estimation in transformation models with parametric regression in the cases where either the transformation or the error distribution or both are modeled nonparametrically. 
Linton, Sperlich and Van Keilegom (2008) consider a parametric class of transformations, while the error distribution is estimated nonparametrically and the regression function is assumed to be additive. The aim of the transformation is to induce independence of the covariate and the error. Asymptotic normality of a profile likelihood estimator for the transformation parameter is proved. 
Heuchenne, Samb and Van Keilegom (2014) consider a residual based empirical distribution function in the same model in order to estimate the error distribution. 

The aim of our paper is twofold.  On one hand we generalize the results of Linton {\it et al.} (2008) by allowing heteroscedasticity. 
To this end in a parametric class of transformations we seek the one that leads to a nonparametric location-scale model of the form
\begin{equation}\label{mod}
\Lambda(Y)=m(X)+\sigma(X)\varepsilon, \mbox{ where } \varepsilon \perp X \mbox{ with } E[\varepsilon]=0, \mbox{Var}(\varepsilon)=1,
\end{equation}
where $\Lambda$ denotes the transformation.
The regression function $m$ and variance function $\sigma^2$ are modeled fully nonparametrically, but analogous results can be obtained for semiparametric modeling. 
We estimate the transformation parameter by a profile-likelihood approach and prove asymptotic normality of the estimator. We investigate the performance of the estimator in a simulation study. 
Note that in the context of parametric regression, Zhou {\it et al.} (2009) and Khan {\it et al.} (2011) considered heteroscedastic transformation models.

On the other hand for the first time in the literature a test for model validity in the context of transformation models with parametric class of transformations and non- (or semi-)parametric regression function is proposed. 
Mu and He (2007) consider estimation procedures in a transformation model with linear quantile regression function and also suggest a test for model validity.
In the general heteroscedastic case we suggest tests for the hypothesis of existence of some transformation $\Lambda$ in the considered parametric class such that the data fulfill model (\ref{mod}). 
The results can readily be modified to test whether such a model can hold with $\sigma\equiv 1$, i.\,e.\ a homoscedastic transformation model.
Our test statistics are based on the difference between the estimated joint distribution of covariables and errors and the product of the marginal distributions. 
A similar approach was used to test for validity of a location-scale model (without transformation) by Einmahl and Van Keilegom (2008). 
However, the estimation of the unknown transformation vastly complicates the theoretical derivations.  
We show weak convergence of the estimated empirical process to a centered Gaussian process under the null hypothesis of model validity. 
As a by-product we obtain an expansion for the residual-based empirical distribution function that generalizes results by Heuchenne {\it et al}.\ (2014).  
Moreover, we discuss consistency of the proposed tests and demonstrate the finite sample properties of a bootstrap version of Kolmogorov-Smirnov
and Cram\'{e}r von Mises tests in a simulation study. 

The rest of the paper is organized as follows. 
In Section 2 we define the profile likelihood estimator for the transformation parameter and show asymptotic normality. 
We further discuss estimation of the regression and variance function by local polynomial estimators, and the estimation of the error distribution. 
In Section 3 we consider the problem of testing for existence of a transformation in the considered class that leads to a location-scale model. We derive an expansion for the estimator of the joint distribution of covariates and errors. 
Under the null hypothesis we show weak convergence of the process given by the difference of the estimated joint distribution and the product of the marginals. 
Consistency of the testing procedures and modifications for the homoscedastic model are discussed.
Additionally, we describe bootstrap versions of the hypothesis tests. 
In Section 4, we also present simulations to demonstrate finite sample properties of the profile likelihood estimator for the transformation parameter as well as the hypothesis tests. 
All regularity conditions and proofs are collected in Appendices A, B and C. 
\section{Estimation of the model} \label{estimation}
Let $L=\{\Lambda_\vartheta\mid \vartheta\in\Theta\}$  be some parametric class of differentiable and strictly increasing transformations, and let $\Theta$ be some nonempty subset of $\mathbb{R}^k$. In this section we assume that there exists some unique $\vartheta_0\in\Theta$ such that
$$\frac{\Lambda_{\vartheta_0}(Y)-E[\Lambda_{\vartheta_0}(Y)|X]}{(\Var(\Lambda_{\vartheta_0}(Y)|X))^{1/2}} \perp X.$$
Then the covariate and transformed response can be modeled by a nonparametric location-scale model, i.\,e.\
\begin{eqnarray} \label{model}
&& \Lambda_{\vartheta_0}(Y)=m(X)+\sigma(X)\eps,\quad \eps\perp X,
\end{eqnarray}
where $m(x)=E[\Lambda_{\vartheta_0}(Y)|X=x]$ and $\sigma^2(x)=\Var(\Lambda_{\vartheta_0}(Y)|X=x)$.

\subsection{Estimation of the transformation parameter}\label{est-trafo}

To estimate the transformation parameter $\vartheta_0$ we will use a profile likelihood approach.  This type of approach has also been used by Linton {\it et al}.\ (2008) in the context of homoscedastic transformation models.  We will extend their method to the current setup with heteroscedastic errors. 

For $\vartheta\in\Theta$, let $m_\vartheta(x)=E[\Lambda_{\vartheta}(Y)|X=x]$, $\sigma_\vartheta^2(x)=\Var[\Lambda_\vartheta(Y)|X=x]$, and
$$\eps(\vartheta)=\frac{\Lambda_\vartheta(Y)-m_\vartheta(X)}{\sigma_\vartheta(X)}.$$
Also, let $F_{\eps(\vartheta)}(y)=P(\eps(\vartheta)\leq y)$ denote the marginal distribution function of the errors and let $f_{\eps(\vartheta)}(y)$ be the corresponding probability density function.  We use the abbreviated notations $\Lambda=\Lambda_{\vartheta_0}$, $\eps=\eps(\vartheta_0)$, $m=m_{\vartheta_0}$, $\sigma^2=\sigma_{\vartheta_0}^2$, $F_\eps=F_{\eps(\vartheta_0)}$ and $f_\eps=f_{\eps(\vartheta_0)}$.

Then, the conditional distribution $F_{Y|X}(\cdot|x)$ of $Y$ given $X=x$ can be written as
$$ F_{Y|X}(y|x) = F_\eps \Big(\frac{\Lambda(y)-m(x)}{\sigma(x)}\Big), $$
and hence the conditional density $f_{Y|X}(\cdot|x)$ of $Y$ given $X=x$ equals
$$ f_{Y|X}(y|x) = f_\eps \Big(\frac{\Lambda(y)-m(x)}{\sigma(x)}\Big) \frac{\Lambda^\prime(y)}{\sigma(x)}. $$

Assume we have independent observations $(X_i,Y_i)$, $i=1,\dots,n$, from the same distribution as $(X,Y)$ and let $\eps_i=\eps_i(\vartheta_0)$, $i=1,\dots,n$.   Then, for an arbitrary value $\vartheta \in \Theta$, the log-likelihood can be written as 
\begin{eqnarray} \label{loglik}
L_\vartheta=\sumi \Big\{\log f_{\eps(\vartheta)} \Big(\frac{\Lambda_\vartheta(Y_i)-m_\vartheta(X_i)}{\sigma_\vartheta(X_i)}\Big) + \log \Lambda_\vartheta^\prime(Y_i) - \log \sigma_\vartheta(X_i) \Big\}. 
\end{eqnarray}

In order to maximize this log-likelihood with respect to $\vartheta$, we first need to replace the unknown functions $f_{\eps(\vartheta)}$, $m_\vartheta$ and $\sigma_{\vartheta}$ by suitable estimators.  For each $\vartheta\in\Theta$ we estimate $m_\vartheta(x)$ by a local polynomial estimator based on $(X_i,\Lambda_\vartheta(Y_i))$, $i=1,\dots,n$.  To this end denote the components of $X_i$ by $(X_{i1},\ldots,X_{id})$ ($i=1,\ldots,n$) and let $x=(x_1,\ldots,x_d)$. Let $\hat m_\vartheta(x) = \hat\beta_0$, where $\hat \beta_0$ is the first component of the vector $\hat \beta$, which is the solution of the local minimization problem
\begin{eqnarray} \label{hm}
\min_\beta \sumi \Big\{\Lambda_\vartheta(Y_i) -  {P_i}(\beta,x,p)\Big\}^2 K_h(X_i-x).
\end{eqnarray}
Here, $ {P_i}(\beta,x,p)$ is a polynomial of order $p$ built up with all $0 \le k \le p$ products of {factors} of the form $X_{ij}-x_j$ ($j=1,\ldots,d$).  The vector $\beta$ is the vector consisting of all coefficients of this polynomial. Here, for $u=(u_1,\ldots,u_d) \in \mathbb{R}^d$, $K(u) = \prod_{j=1}^d k(u_j)$ is a $d$-dimensional product kernel, $k$ is a univariate kernel function, $h = (h_1,\ldots,h_d)$ is a $d$-dimensional bandwidth vector converging to zero when $n$ tends to infinity, and $K_h(u) = \prod_{j=1}^d k(u_j/h_j)/h_j$. 

Analogously, for each $\vartheta\in\Theta$ let  $\hat s_\vartheta$ denote a local polynomial estimator based on $(X_i,\Lambda_\vartheta(Y_i)^2)$, $i=1,\dots,n$, and define the variance function estimator as $\hat\sigma^2_\vartheta=\hat s_\vartheta-\hat m_\vartheta^2$. Note that this estimator has similar properties as a local polynomial estimator based on $(X_i,(\Lambda_\vartheta(Y_i)-\hat m_\vartheta(X_i))^2)$, $i=1,\dots,n$.   

Finally, let $\hat \eps_i(\vartheta) = (\Lambda_\vartheta(Y_i)-\hat m_\vartheta(X_i))/\hat \sigma_\vartheta(X_i)$ and define
$$ \hat f_{\hat\eps(\vartheta)}(y) = \frac{1}{n} \sumi \ell_g \big(\hat \eps_i(\vartheta)-y\big), $$
where $\ell$ and $g$ are a kernel function and a bandwidth sequence, possibly different from the kernel $k$ and the bandwidth $h$ that were used to estimate the regression and variance function.

Next, we plug in the estimators $\hat m_\vartheta$, $\hat \sigma_\vartheta$ and $\hat f_{\hat\eps(\vartheta)}$ into the log-likelihood given in (\ref{loglik}) and obtain the following profile likelihood estimator of $\vartheta$:
\begin{eqnarray} \label{hattheta}
\hat \vartheta = \mbox{argmax}_{\vartheta \in \Theta} \sumi \Big\{\log \hat f_{\hat\eps(\vartheta)} \Big(\frac{\Lambda_\vartheta(Y_i) - \hat m_\vartheta(X_i)}{\hat \sigma_\vartheta(X_i)}\Big) + \log \Lambda_\vartheta^\prime(Y_i) - \log \hat \sigma_\vartheta(X_i) \Big\}. 
\end{eqnarray}

In order to obtain an asymptotic i.i.d.\ representation and the asymptotic normality of the estimator $\hat \vartheta$, we need to introduce a number of notations.   For any function $h_\vartheta$ we denote by $\dot h_\vartheta = \nabla_\vartheta h_\vartheta$ the vector of partial derivatives of $h_\vartheta$ with respect to the components of $\vartheta$.  Let
$$ G_n (\vartheta) = \frac{1}{n} \sumi g_\vartheta(X_i,Y_i) $$
be the derivative of the log-likelihood given in (\ref{loglik}) (divided by $n$) with respect to $\vartheta$, where
\begin{eqnarray*}
g_\vartheta(X_i,Y_i) &\m = &\m \frac{f^\prime_{\eps(\vartheta)}(\eps_i(\vartheta))}{f_{\eps(\vartheta)}(\eps_i(\vartheta))} \Big[\frac{\dot \Lambda_\vartheta(Y_i)-\dot m_\vartheta(X_i)}{\sigma_\vartheta(X_i)} - \{\Lambda_\vartheta(Y_i)-m_\vartheta(X_i)\} \frac{\dot \sigma_\vartheta(X_i)}{\sigma_\vartheta^2(X_i)} \Big] \\
&\m &\m + \frac{\dot f_{\eps(\vartheta)}(\eps_i(\vartheta))}{f_{\eps(\vartheta)}(\eps_i(\vartheta))} + \frac{\dot \Lambda^\prime_\vartheta(Y_i)}{\Lambda^\prime_\vartheta(Y_i)} - \frac{\dot \sigma_\vartheta(X_i)}{\sigma_\vartheta(X_i)}.
\end{eqnarray*}
Then $G_n(\vartheta)$ converges in probability to  $G(\vartheta) = E[g_\vartheta(X,Y)]$.  We assume that $\vartheta_0$ is the unique zero of $G$ (see assumption (a7) in appendix A). 
The next theorem states the asymptotic normality of the estimator $\hat \vartheta$.  The result shows that the variance of the estimator is the same as in the case where the nonparametric functions $m_\vartheta(x)$, $\sigma_\vartheta(x)$ and $f_{\eps(\vartheta)}(y)$ and their derivatives with respect to $\vartheta$ and $y$ would be known, which is quite remarkable.   The regularity conditions under which this result is valid are given in appendix A.  

\begin{theorem} \label{asnotheta}
Assume (a1)--(a7) in Appendix A.  Then, 
$$ \hat \vartheta - \vartheta_0 = - \Gamma^{-1} \frac{1}{n} \sumi g_{\vartheta_0}(X_i,Y_i) + o_P(n^{-1/2}), $$
and
$$ n^{1/2} \big(\hat \vartheta - \vartheta_0\big) \stackrel{d}{\rightarrow} N \big(0, \Sigma \big), $$
where \rm $\Sigma = \Gamma^{-1} \mbox{Var}[g_{\vartheta_0}(X,Y)] \Gamma^{-1}$ \it and $\Gamma = \nabla_\vartheta G(\vartheta)^\top|_{\vartheta=\vartheta_0}$. 
\end{theorem}

The proof of this result can be found in Appendix B.

\subsection{Estimation of regression and variance functions}

Once the transformation parameter vector $\vartheta_0$ is estimated, we can go back to the estimation of the regression function $m(x)$ and the variance function $\sigma^2(x)$.  Define 
$$ \hat m(x) = \hat m_{\hat \vartheta}(x) \hspace*{1cm} \mbox{and} \hspace*{1cm} \hat \sigma^2(x) = \hat \sigma_{\hat \vartheta}^2(x). $$  
Under regularity conditions the estimation of $\vartheta_0$ has no influence on the asymptotic distribution of the centered and scaled estimators $(nh^d)^{1/2}(\hat m(x)-E[\hat m(x)])$ and $(nh^d)^{1/2}(\hat \sigma^2(x)-E[\hat \sigma^2(x)])$, since $\hat \vartheta$ has a parametric rate of convergence.  Therefore, the estimators behave asymptotically as if the true $\vartheta_0$ would be known. Note, however, that the pre-estimation of $\vartheta_0$ influences the asymptotic distribution of the test statistic in Section \ref{testing} because the integrals $\int (\hat m_{\vartheta_0}-m)/\sigma \, dF_X$ and $\int (\hat m_{\hat \vartheta}-\hat m_{\vartheta_0})/\sigma \, dF_X$ have the same $n^{1/2}$-rate of convergence (see terms $B_n$ and $C_n$ in the proof of Theorem \ref{theo1}) and a similar statement holds for the variance estimator.

\subsection{Estimation of the error distribution}\label{est-error-distr}

The last unknown component of our heteroscedastic transformation model (\ref{model}) is the distribution $F_\eps$ of the error term.   Define the residuals as 
\begin{eqnarray*}
\hat\eps_i=\hat\eps_i(\hat\vartheta)=\frac{\Lambda_{\hat\vartheta}(Y_i)-\hat m(X_i)}{\hat\sigma(X_i)}. 
\end{eqnarray*}

The error distribution $F_\eps(y)$ can now be estimated by the empirical distribution function of the $\hat \eps_i$'s:
$$ \hat F_{\hat\eps}(y) = \frac{1}{n} \sum_{i=1}^n I\{\hat\eps_i\leq y\}, $$
where $I$ denotes the indicator function.  We postpone the study of the asymptotic properties of this estimator to the next section.  In fact, in Section \ref{testing} we will study an estimator of the joint distribution of $X$ and $\eps$, which includes the estimator $\hat F_{\hat\eps}(y)$ as a special case.

\section{Testing the validity of the model} \label{testing}

In this section we develop tests for validity of a heteroscedastic semiparametric transformation model. Let again $L=\{\Lambda_\vartheta\mid \vartheta\in\Theta\}$  be some parametric class of  transformations, $\Theta$ some nonempty subset of $\mathbb{R}^k$. Our aim is to test the null hypothesis
\begin{eqnarray} \label{heter}
&& H_0:\exists\vartheta\in\Theta \mbox{ such that } \frac{\Lambda_\vartheta(Y)-E[\Lambda_\vartheta(Y)|X]}{(\Var(\Lambda_\vartheta(Y)|X))^{1/2}} \perp X.
\end{eqnarray}
If the null hypothesis is valid then there exists some transformation $\Lambda_{\vartheta_0}\in L$ with which one obtains a nonparametric location-scale model as in (\ref{model}). 
Note that we want to test the appropriateness of the parametric family of transformations. So, our test is a goodness-of-fit test for the chosen parametric family.  We do not test whether data is from a transformation model or not.  If we reject $H_0$ it could be that the data is from a transformation model but that the true transformation does not belong to the family $L$ under our consideration. 

\subsection{The test statistics and asymptotic distributions under $H_0$}\label{4.1}

Let $\hat \vartheta$ be some estimator for the true parameter $\vartheta_0$ under $H_0$ such that a linear expansion
\begin{eqnarray}\label{lin-exp-theta}
\hat\vartheta-\vartheta_0 &=& \frac 1n \sum_{i=1}^n g_{\vartheta_0}(X_i,Y_i)+o_P \left(\frac{1}{\sqrt{n}} \right)
\end{eqnarray}
is valid under $H_0$, where $E[g_{\vartheta_0}(X_i,Y_i)]=0$, $E[\|g_{\vartheta_0}(X_i,Y_i)\|^2]<\infty$. We have shown in Theorem \ref{asnotheta} that such an expansion is valid for the profile likelihood estimator under some regularity conditions. 
Now denote by $\hat F_{X,\hat\eps}$ the joint empirical distribution function of covariates and residuals,  i.\,e.\
$$\hat F_{X,\hat\eps}(x,y)=\frac{1}{n}\sum_{i=1}^n I\{X_i\leq x,~\hat\eps_i\leq y\},$$
where $\leq$ for vectors is meant componentwise. We consider test statistics based on the estimated independence empirical process
\begin{eqnarray}\label{Sn}
S_n=\sqrt{n}(\hat F_{X,\hat\eps}-\hat F_{X}\hat F_{\hat\eps})
\end{eqnarray}
where $\hat F_{X}(x)=n^{-1}\sum_{i=1}^n I\{X_i\leq x\}$ and $\hat F_{\hat\eps}(y)=n^{-1}\sum_{i=1}^n I\{\hat\eps_i\leq y\}$.

\bth\label{theo1} Assume (a1), (a2) and (A1)--(A8) from appendix A.  Then, under $H_0$, we have the asymptotic expansion:
\begin{eqnarray*}
\hat F_{X,\hat\eps}(x,y)&=&\frac{1}{n}\sum_{i=1}^n \Big(I\{X_i\leq x\}\Big(I\{\eps_i\leq y\}+f_\eps(y)(\eps_i+\frac{y}{2}(\eps_i^2-1))\Big)\\
&&{}+E\Big[\nabla_\vartheta F_{\eps(\vartheta)|X}(y|X)|_{\vartheta=\vartheta_0}I\{X\leq x\}\Big]^\top  g_{\vartheta_0}(X_i,Y_i)\Big)
+o_P(n^{-1/2})
\end{eqnarray*}
uniformly with respect to $x\in R_X, y\in\mathbb{R}$. 
\eth

The proof is given in appendix B.  From the theorem one directly obtains the following result for the residual based empirical distribution function defined in Section \ref{est-error-distr}.

\bco\label{cor1}  Under the assumptions of Theorem \ref{theo1}, we have the asymptotic expansion:
\begin{eqnarray*}
\hat F_{\hat\eps}(y)&=&\frac{1}{n}\sum_{i=1}^n \Big(I\{\eps_i\leq y\}+f_\eps(y)(\eps_i+\frac{y}{2}(\eps_i^2-1))\Big)\\
&&{}+E\Big[\nabla_\vartheta F_{\eps(\vartheta)|X}(y|X)|_{\vartheta=\vartheta_0}\Big]^\top  g_{\vartheta_0}(X_i,Y_i)\Big)
+o_P(n^{-1/2})
\end{eqnarray*}
uniformly with respect to $y\in\mathbb{R}$. The process $\sqrt{n}(\hat F_{\hat\eps}-F_\eps)$ converges weakly in $\ell^\infty(\mathbb{R})$ to a centered Gaussian process.
\eco

This corollary generalizes the main results by Heuchenne {\it et al.} (2014) who consider estimation of the error distribution in a homoscedastic transformation model. 
The  asymptotic expansion directly follows from Theorem \ref{theo1}. The proof of weak convergence is analogous to the proof of Corollary \ref{cor1} below and thus omitted.

Using that the dominating term in this expansion has expectation $F_\eps(y)$ and applying that $\hat F_X=F_X+O_p(n^{-1/2})$ one straightforwardly obtains the following expansion for the process $S_n$ defined in (\ref{Sn}):
\begin{eqnarray}\label{sncor}
S_n(x,y)&=& \frac{1}{\sqrt{n}}\sum_{i=1}^n \psi_{x,y,\vartheta_0}(X_i,Y_i)+o_P(1)
\end{eqnarray}
uniformly with respect to $x\in R_X, y\in\mathbb{R}$, where
\begin{eqnarray*}
\psi_{x,y,\vartheta_0}(X_i,Y_i)
&=&
\Big(I\{X_i\leq x\}-F_X(x)\Big)\Big(I\{\eps_i\leq y\}-F_\eps(y)+f_\eps(y)(\eps_i+\frac{y}{2}(\eps_i^2-1))\Big)\\
&&{}+E\Big[\nabla_\vartheta F_{\eps(\vartheta)|X}(y|X)|_{\vartheta=\vartheta_0}\Big(I\{X\leq x\}-F_X(x)\Big)\Big]^\top  g_{\vartheta_0}(X_i,Y_i).
\end{eqnarray*}

\bco\label{cor1}  Under the assumptions of Theorem \ref{theo1}, the process $S_n$ converges weakly in $\ell^\infty(R_X\times\mathbb{R})$ to a centered Gaussian process $S$ with covariance $\Cov(S(x,y),S(u,z))=E[ \psi_{x,y,\vartheta_0}(X,Y)\psi_{u,z,\vartheta_0}(X,Y)]$.
\eco

The proof is given in appendix B. 
Let $\Psi$ denote some continuous functional from $\ell^\infty(R_X\times\mathbb{R})$ to $\mathbb{R}$, e.\,g.\ $\Psi(s)=\sup_{x,y}|s(x,y)|$ for a Kolmogorov-Smirnov test. Then we reject $H_0$ with nominal level $\alpha$ if $T_n=\Psi(S_n)$ exceeds a critical value $c_\alpha$. A bootstrap approximation of $c_\alpha$ is given in Section \ref{bootstrap}.  

\subsection{Bootstrap approximation of the critical value} \label{bootstrap}
Since the asymptotic distributions of the test statistics depend in a complicated way on unknown quantities, we suggest to apply a bootstrap procedure to approximate the critical values. To this end let $\eta_1^*,\dots,\eta_n^*$ be drawn with replacement from standardized residuals $\tilde\eps_1,\dots,\tilde\eps_n$, where
\begin{equation} \label{err}
 \tilde \eps_i=\frac{\hat\eps_i-n^{-1}\sum_{k=1}^n \hat\eps_k}{(n^{-1}\sum_{j=1}^n (\hat\eps_j-n^{-1}\sum_{k=1}^n \hat\eps_k))^{1/2}},\quad i=1,\dots,n.
\end{equation}
Let further $\xi_1,\dots,\xi_n$ denote independent standard normally distributed random variables, independent of the original sample $\Y_n=\{(X_1,Y_1),\dots,(X_n,Y_n)\}$, and let $a_n$ be some positive smoothing parameter.  Define bootstrap errors as $\eps_i^*=\eta_i^*+a_n\xi_i$. Note that methods based on residual empirical processes require smoothing of the bootstrap errors, cf.\ Neumeyer (2009b), among others. 
It is easily seen that, conditionally on $\Y_n$, $\eps_i^*$ has a smooth distribution function
$$\tilde F_{\hat\eps}(y)=\frac{1}{n}\sum_{j=1}^n \Phi(\frac{y-\tilde{\eps_j}}{a_n}),$$
where $\Phi$ denotes the standard normal distribution function. 

Now generate $X_i^*$ from $\hat F_X$ and define 
\begin{equation} \label{bmodel}
Y_i^*=\Lambda_{\hat\vartheta}^{-1}(Z_i^*),\mbox{ where }Z_i^*=\hat m(X_i^*)+\hat\sigma(X_i^*)\eps_i^*,\quad i=1,\dots,n. 
\end{equation}
The bootstrap sample is $(X_i^*,Y_i^*)$, $i=1,\dots,n$, and fulfills $H_0$ by construction.  To see this let $E_n^*$ and $\Var_n^*$ denote the expectation and variance with respect to the conditional distribution $P(\cdot\mid \Y_n)$. Then $E_n^*[\eps_i^*\mid X_i^*]\equiv 0$ and $\Var_n^*(\eps_i^*\mid X_i^*)\equiv 1+a_n^2$ and thus
$$\frac{\Lambda_{\hat\vartheta}(Y_i^*)-E_n^*[\Lambda_{\hat \vartheta}(Y_i^*)|X_i^*]}{(\Var_n^*(\Lambda_{\hat \vartheta}(Y_i^*)|X_i^*))^{1/2}}=\frac{\eps_i^*}{(1+a_n^2)^{1/2}} \perp X_i^*$$
(given $\Y_n$).
Let $T_n$ denote the test statistic based on the original sample and let $T_n^*$ be the one based on the bootstrap sample. Then $H_0$ is rejected whenever $T_n>c_{n,\alpha}$, where
$P(T_n^*>c_{n,\alpha}\mid \Y_n)=1-\alpha$.
The critical value $c_{n,\alpha}$ is estimated by the $\lfloor B(1-\alpha)\rfloor$-largest bootstrap test statistic obtained from $B$ replications of the bootstrap data generation. 

\subsection{Remarks on consistency of the proposed tests}

We consider the hypothesis test developed in Section \ref{4.1} when using the profile likelihood estimator $\hat\vartheta$ suggested in Section \ref{est-trafo}.
With the notations used before let
$$p_\vartheta(y|x)= f_{\eps(\vartheta)}\Big(\frac{\Lambda_\vartheta(y)-m_\vartheta(x)}{\sigma_\vartheta(x)}\Big)\frac{\Lambda_\vartheta'(y)}{\sigma_\vartheta(x)}.$$
Note that $p_\vartheta$ is a conditional density, and a consistent estimator (under mild regularity conditions) of the log-likelihood
$$L_\vartheta=\log\Big(\prod_{i=1}^n p_\vartheta(Y_i|X_i)\Big)$$
is maximized in order to obtain the profile likelihood estimator of the transformation parameter $\vartheta\in\Theta$ (see (\ref{loglik})). Now, consider the alternative $H_1$, which states that there exists no parameter $\vartheta\in\Theta$ such that $p_\vartheta(\cdot|x)$ is the conditional density of $Y$, given $X=x$. Then $L_\vartheta/n$ estimates the expectation
$$E[\log p_\vartheta(Y_i|X_i)]=\int\int (\log p_\vartheta(y|x))f_{Y|X}(y|x)\,dy dF_X(x)$$
and thus $\hat\vartheta$ estimates the value $\vartheta_1\in\Theta$ which minimizes the expected Kullback-Leibler divergence of the conditional densities $f_{Y|X}$ and $p_\vartheta$, i.\,e.\
$$\int\int \Big(\log \frac{f_{Y|X}(y|x)}{p_\vartheta(y|x)}\Big)f_{Y|X}(y|x)\,dy dF_X(x).$$
Thus $\hat F_{X,\hat\eps}$ as defined in section \ref{4.1} estimates the joint distribution of $X$ and 
$\eps(\vartheta_1)=(\Lambda_{\vartheta_1}(Y)-E[\Lambda_{\vartheta_1}(Y)|X])/(\Var(\Lambda_{\vartheta_1}(Y)|X))^{1/2}$. 
Since under $H_1$ the distribution of $\eps(\vartheta_1)$ depends on  $X$, it follows that, e.\,g., a Kolmogorov-Smirnov test statistic $T_n=\sup_{x,y}|S_n(x,y)|$ converges to infinity. Thus any test that rejects $H_0$ whenever $T_n$ exceeds some constant $c_\alpha$ is consistent. 

\subsection{The homoscedastic transformation model}\label{hom}

Let independent copies of $(X,Y)$ be observed and a parametric class of transformations $\{\Lambda_\vartheta\mid\vartheta\in\Theta\}$ be given. 
Then tests for the null hypothesis
\begin{eqnarray} \label{htest}
H_0:\exists\vartheta\in\Theta \mbox{ such that } \Lambda_\vartheta(Y)-E[\Lambda_\vartheta(Y)|X] \perp X
\end{eqnarray}
are also of interest. The validity of the null hypothesis means that a nonparametric location model
$$\Lambda_{\vartheta_0}(Y)=m(X)+\eps,\quad \eps\perp X$$
with $m(x)=E[\Lambda_{\vartheta_0}(Y)|X=x]$ describes the data for some $\vartheta_0\in\Theta$. Tests for model validity can be derived similarly as in the heteroscedastic case in an obvious manner. An estimator for the transformation parameter analogous to Linton {\it et al}.\ (2008) can be applied where the additive regression estimator is replaced by a purely nonparametric local polynomial estimator. The residuals are then defined as
$\hat\eps=\Lambda_{\hat\vartheta}(Y)-\hat m_{\hat\vartheta}(X)$.
 Under slightly weaker assumptions than those stated in Appendix A, similar asymptotic results to those in Section \ref{4.1} can be derived. Additionally, we can use the simplification of the bootstrap in Section \ref{bootstrap} to implement the test for the validity of (\ref{htest}) replacing $\tilde \eps_i$ in (\ref{err}) with $\tilde \eps_i = \hat \eps_i - n^{-1} \sum_{k=1}^n \hat \eps_k$, and $Z_i^*$ in (\ref{bmodel}) with $Z_i^* = \hat m(X_i^*) + \eps_i^*$.

\section{Numerical simulations}

In this section, we carry out three different simulation studies. Firstly, we illustrate the finite sample performance of the estimator $\hat \vartheta$ of the transformation parameter  in (\ref{hattheta}). Secondly, we study the performance of the proposed test for checking homoscedasticity under some transformation when it is implemented via the bootstrap described in Section \ref{hom}. Finally, we verify how well the test in Section \ref{4.1} is able to test the assumption of a heteroscedastic transformation structure, when the true model gradually deviates from a heteroscedastic transformation model. 

Throughout all simulations, we consider the Yeo-Johnson family of transformations:
$$\Lambda_\vartheta(y)=\left\{
\begin{array}{lr} \{ (y+1)^\vartheta-1 \}/{\vartheta} & y\geq 0,\vartheta\neq 0 \,\\
\log(y+1) & y\geq 0,\vartheta=0\,\\
-\{(-y+1)^{2-\vartheta}-1\}/(2-\vartheta) & y< 0,\vartheta\neq 2 \,\\
-\log(-y+1) & y<0,\vartheta=2
\end{array}\right.,$$
which was proposed by Yeo and Johnson (2000) as a generalization of the Box-Cox family of transformations. 
Concerning the estimation of the transformation parameter, we use the normal kernel whenever a kernel function is necessary. To estimate $m(\cdot)$ and $\sigma(\cdot)$, we use the local linear estimator ($p=1$) and the bandwidth is chosen by the direct plug-in methodology described by Ruppert, Sheather and Wand (1995). For estimation of $f_{\eps(\vartheta)}(\cdot)$, we use the bandwidth obtained from the method of Sheather and Jones (1991). With regard to the test statistics, we consider the Kolmogorov-Smirnov and Cram\'{e}r-von Mises test statistics:    
\begin{eqnarray}
T_{n,KS} &=& \sqrt{n} \sup_{x,y} | \hat F_{X,\hat \eps}(x,y) - \hat F_X(x) \hat F_{\hat \eps}(y)|;
\\ T_{n,CM} &=& n \int \int (\hat F_{X,\hat \eps}(x,y) - \hat F_X(x) \hat F_{\hat \eps}(y))^2 d\hat F_X(x) d\hat F_{\hat \eps}(y). 
\end{eqnarray}
To find the critical value for the proposed tests, we use 200 bootstrap replications for each sample. For the smooth bootstrap described in Section \ref{bootstrap}, we set $a_n$ to $0.5n^{-1/4}$ as in Neumeyer (2009b).  

\subsection{Estimation of heteroscedastic transformation parameter}
To see how the estimator $\hat \vartheta$ in (\ref{hattheta}) works in practice, we generate data from the following heteroscedastic transformation model:     
\begin{eqnarray} \label{htm1}
\Lambda_{\vartheta_0=0}(Y_i) = m(X_i)+\sigma(X_i) \varepsilon_i,~i=1,\cdots,n,
\end{eqnarray}
where $m(x)=\exp(x)+1.5$, $\sigma(x) = 1+a(x-1)$, $X \sim U[0,1]$, $\varepsilon \sim N(0,1^2)$ and $X \perp \varepsilon$. For various values of $a$ and $n$, we calculate $\hat \vartheta$ from 200 samples of size $n=100,200$ and $400$, and compute 
$$ {\rm MEAN}=\frac 1 {200} \sum_{j=1}^{200}\hat \vartheta^{(j)}~{\rm and}~{\rm MSE}=\frac 1 {200} \sum_{j=1}^{200} (\hat \vartheta^{(j)}-\vartheta_0)^2,$$
where $\hat \vartheta^{(j)}$ is the estimate of $\vartheta_0$ from the $j$th sample. The results are given in Table \ref{table1}. For various values of $a$, we observe that both the bias and the mean squared error of the estimator decrease as the sample size increases, which suggests the consistency of the estimator.   

\begin{table}[h]
\begin{center}
\begin{tabular}{lcccccccc}
\hline
 & \multicolumn{2}{c}{$n=100$} & & \multicolumn{2}{c}{$n=200$} & &\multicolumn{2}{c}{$n=400$}\\ 
 \cline{2-3} \cline{5-6} \cline{8-9}  
 & MEAN & MSE & &MEAN & MSE & & MEAN & MSE \\
\hline
$a=0.5$ & 0.085 & 0.198 & & 0.035 & 0.117 & & 0.026 & 0.062 \\
$a=0.75$ & 0.077 & 0.200 & & 0.048 & 0.090 & & 0.008 & 0.053 \\
$a=1$ & 0.056 & 0.228 & & 0.074 & 0.121 & & -0.009 & 0.066 \\
\hline 
\end{tabular}
\caption{The bias and mean squared error of the estimator $\hat \vartheta$ for $n=100,~200$ and $400$. } \label{table1}
\end{center}
\end{table}

\subsection{Testing for homoscedastic transformation models}

To verify the performance of the test proposed in Section \ref{hom} regarding the assumption of a homoscedastic transformation model, we reuse model (\ref{htm1}). Note that the degree of heteroscedasticity decreases as the value of $a$ gets closer to 0 and model (\ref{htm1}) becomes a homoscedastic transformation model when $a=0$, which satisfies the null hypothesis (\ref{htest}). We investigate how the test behaves as the value of $a$ increases from 0 to 1. 

\begin{figure}[h!]
\begin{center}
\includegraphics[width=13cm]{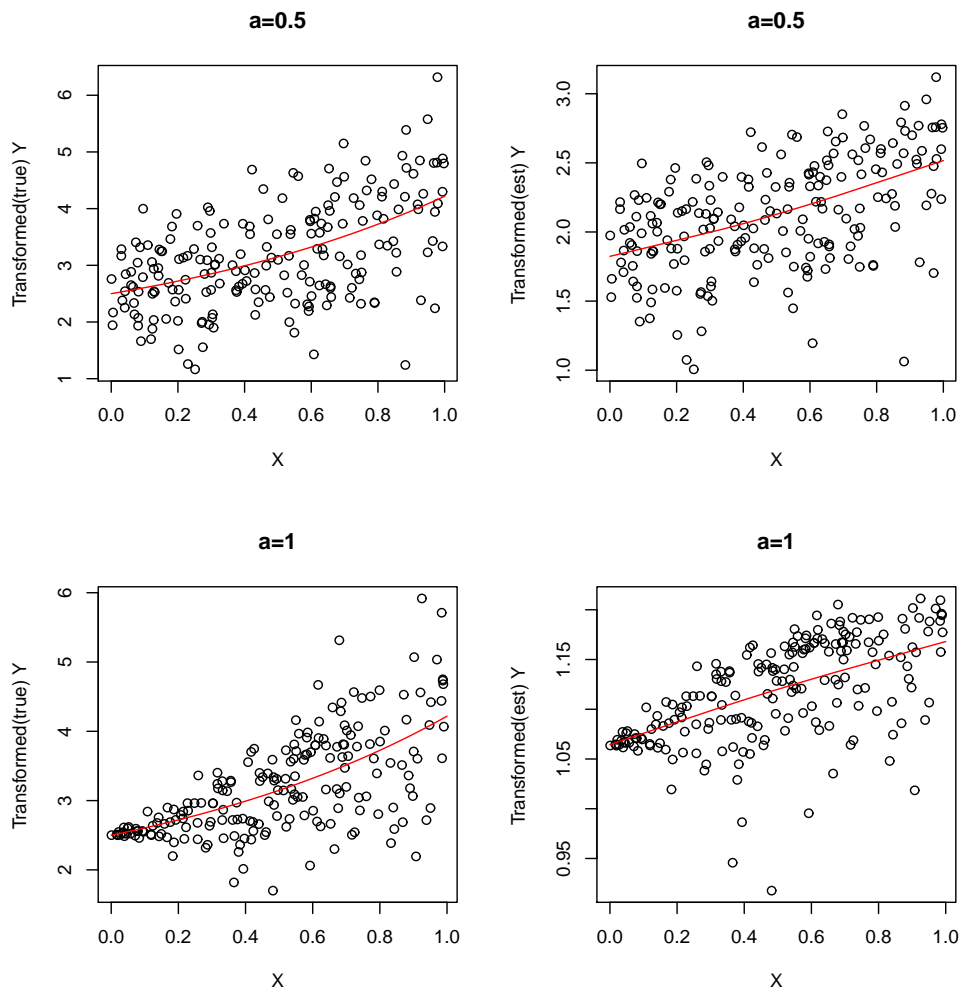}
\end{center}
\caption{Plot of $\Lambda_{\vartheta=\vartheta_0}(Y_i)$ versus $X_i$ (left panel), and $\Lambda_{\vartheta=\hat \vartheta}(Y_i)$ versus $X_i$ (right panel), when $a=0.5$ (upper panel) and $a=1$ (lower panel).  The curves $m_{\vartheta_0}(\cdot)$ (left) and $m_{\hat\vartheta}(\cdot)$ (right) are indicated in red.} \label{reason}
\end{figure}

Table \ref{table2} shows the results for the test implemented via the bootstrap described in Section \ref{hom}. We see that the size of the test is somewhat too low, but the power grows to one as the parameter $a$ measuring the degree of heteroscedasticity gets larger. One notable feature of the results is that the power stays flat until the degree of heteroscedasticity reaches a certain level and then suddenly starts to increase. To explain this peculiar behavior, we show in Figure \ref{reason} four plots using data of size $n=200$ from model (\ref{htm1}). 
These plots are given for two values of $a$, and compare the regression function based on the true parameter $\vartheta_0$ with the one based on the estimator $\hat \vartheta$. 

When $a \neq 0$, the estimator $\hat \vartheta$ is not consistent due to the misspecification of the heteroscedastic error structure, and instead targets the pseudo-true parameter $\vartheta^* \neq \vartheta_0$ which maximizes 
\begin{eqnarray}
PL(\vartheta) &=& E (\log f_{\varepsilon_{\vartheta}}(\Lambda_{\vartheta}(Y)-m_{\vartheta}(X)) + \log \Lambda'_{\vartheta}(Y)),
\end{eqnarray}
where $m_{\vartheta}(x) = E(\Lambda_{\vartheta}(Y)|X=x)$ and $\varepsilon_{\vartheta} = \Lambda_{\vartheta}(Y)- m_{\vartheta}(X)$. This pseudo-true parameter has the interpretation that the corresponding homoscedastic model is the best approximation to the true heteroscedastic transformation model. So when the degree of heteroscedasticity is moderate, it is possible that the data look like data coming from a homoscedastic transformation model with transformation parameter $\hat \vartheta$ (see the upper right panel of Figure \ref{reason}). In this case, our test is not able to detect the violation of assumption (\ref{htest}) well, and behaves almost as if the null hypothesis is true. However, when the degree of heteroscedasticity becomes severe, the data cannot be considered anymore to come from a homoscedastic transformation model, and it becomes possible to detect the violation through the dependence between $X$ and $\hat \eps$ (see the right lower panel of Figure \ref{reason}). This feature is different from what was observed in testing for homoscedasticity in regression settings without transformation, such as in Neumeyer (2009a). 
 
\begin{table}[h]
\begin{center}
\begin{tabular}{lccccccccccc}
\hline
 & \multicolumn{5}{c}{$n=100$}& $\phantom{aaa}$ & \multicolumn{5}{c}{$n=200$}\\ \cline{2-6} \cline{8-12}
 & \multicolumn{2}{c}{$\alpha=0.05$} & & \multicolumn{2}{c}{$\alpha=0.1$} & &\multicolumn{2}{c}{$\alpha=0.05$}& & \multicolumn{2}{c}{$\alpha=0.1$}\\ 
 \cline{2-3} \cline{5-6} \cline{8-9} \cline{11-12} 
 & KS & CM & &KS & CM & & KS & CM & & KS & CM\\
\hline
$a=0$ & 0.025 & 0.025 & & 0.060 & 0.065 & & 0.035 & 0.020& & 0.075 & 0.075 \\
$a=0.5$ & 0.070 & 0.085 & & 0.115 & 0.135 & & 0.100 & 0.110 & & 0.145 & 0.200 \\
$a=0.75$ & 0.260 & 0.345 & & 0.345 & 0.420 & & 0.400 & 0.485 & & 0.545 & 0.580 \\
$a=1$ & 0.905 & 0.955 & & 0.970 & 0.980 & & 1.000 & 1.000 & & 1.000 & 1.000 \\ \hline 
\end{tabular}
\caption{The power of the test for verifying the validity of a homoscedastic transformation structure. The power is calculated based on 200 samples. The null hypothesis is satisfied for $a=0$.} \label{table2}
\end{center}
\end{table}

\subsection{Testing for heteroscedastic transformation models}
Finally, we illustrate how the test in Section \ref{4.1} works to verify the assumption of a heteroscedastic transformation structure. For this purpose, we define two new transformation models. Basically, they are the same model as the model (\ref{htm1}), except that the error distribution is defined by \newpage
\noindent \underline{Model A}
\begin{center}
 $(\varepsilon | X=x) \sim \left\{ \begin{array}{ll}
         N(0,1^2) & \mbox{if}~0.5< x \leq 1;\\
        (W-E(W))/\sqrt{Var(W)},~\mbox{where}~W \sim ST(0,1,\alpha,\nu) & \mbox{if}~0\leq x \leq 0.5,\end{array} \right.$
\end{center}
\noindent \underline{Model B}
\begin{center}
($\epsilon | X=x) = \left\{ \begin{array}{ll}
         N(0,1^2) & \mbox{if}~0.5< x \leq 1;\\
        (W-\eta)/\sqrt{2\eta},~\mbox{where}~W \sim \chi^2(\eta)& \mbox{if}~0\leq x \leq 0.5.\end{array} \right.$
\end{center}
Here, $ST(\xi,\Omega,\alpha,\nu)$ is a skew-$t$ distribution with parameters $\xi,\Omega,\alpha$ and $\nu$ defined in Azzalini (2005). The parameter $\alpha$ controls the skewness of the distribution and the paramer $\nu$ controls kurtosis. Additionally, we set $\sigma(x)=x$ (so $a=1$). First, note that as $\nu \rightarrow \infty$ and $\alpha \rightarrow 0$, Model A converges to model (\ref{htm1}) with $\sigma(x)=x$, which satisfies the assumption of a heteroscedastic transformation structure (the same thing happens as $\eta \rightarrow \infty$ in case of Model B). An additional remark regarding these models is that the first and second moments of the conditional error distribution given $X$ coincide with the respective moments under model (\ref{htm1}).  The parameters $\alpha$, $\nu$ and $\eta$ determine how much the model violates assumption (\ref{heter}).  In our simulations, to see how the test performs when the true model gradually deviates from the assumption under the null hypothesis, we investigate the power function as $\nu$ changes from $\infty$ to 2.1 and then as $\alpha$ changes from 0 to 100 for Model A, and as $\eta$ changes from $\infty$ to 2 for Model B. Here, $\nu$ should be greater than 2 and $\eta$ should be equal to or greater than 2 otherwise the distribution of $W$ cannot be standardized due to variance explosion.  

 Similarly to what was observed in the case of homoscedastic transformation models, we observe from Tables \ref{table3} and \ref{table4} that there is a threshold of difference in two component distributions in the error above which we can detect the violation of the assumption, and the power starts to grow beyond the threshold. Further, we observe that compared to Model B, the power of Model A is somewhat lower. The reason can be attributed to the flexibility of the heteroscedastic transformation model. Since they are very flexible models, unless the two component distributions in the error are strikingly different from each other, the generated data look like data coming from a heteroscedastic transformation model with appropriately chosen transformation parameter.   

\begin{table}[h]
\begin{center}
\begin{tabular}{lccccccccccc}
\hline
 & \multicolumn{5}{c}{$n=100$}& $\phantom{aaa}$ & \multicolumn{5}{c}{$n=200$}\\ \cline{2-6} \cline{8-12}
 & \multicolumn{2}{c}{$\alpha=0.05$} & & \multicolumn{2}{c}{$\alpha=0.1$} & &\multicolumn{2}{c}{$\alpha=0.05$}& & \multicolumn{2}{c}{$\alpha=0.1$}\\ 
 \cline{2-3} \cline{5-6} \cline{8-9} \cline{11-12} 
 & KS & CM & &KS & CM & & KS & CM & & KS & CM\\
\hline
$\alpha=100,~~~\nu=2.1$ & 0.370 & 0.445 & & 0.505 & 0.590 & & 0.710 & 0.770 & & 0.795 & 0.850 \\
$\alpha=0,\phantom{00}~~~\nu=2.1$ & 0.105 & 0.140 & & 0.170 & 0.200 & & 0.205 & 0.270 & & 0.325 & 0.360 \\
$\alpha=0,\phantom{00}~~~\nu=5$ & 0.075 & 0.060 & & 0.105 & 0.085 & & 0.060 & 0.060 & & 0.130 & 0.095 \\
$\alpha=0,\phantom{00}~~~\nu=\infty$ & 0.055 & 0.060 & & 0.070 & 0.105 & & 0.080 & 0.070 & & 0.120 & 0.135 \\ \hline  
\end{tabular}
\caption{The power of the test for verifying the validity of a heterocedastic transformation structure from Model A. The power is calculated based on 200 samples. The null hypothesis is satisfied for $\alpha=0$ and $\nu=\infty$.}\label{table3}
\end{center}
\end{table}

\begin{table}[h]
\begin{center}
\begin{tabular}{lccccccccccc}
\hline
 & \multicolumn{5}{c}{$n=100$}& $\phantom{aaa}$ & \multicolumn{5}{c}{$n=200$}\\ \cline{2-6} \cline{8-12}
 & \multicolumn{2}{c}{$\alpha=0.05$} & & \multicolumn{2}{c}{$\alpha=0.1$} & &\multicolumn{2}{c}{$\alpha=0.05$}& & \multicolumn{2}{c}{$\alpha=0.1$}\\ 
 \cline{2-3} \cline{5-6} \cline{8-9} \cline{11-12} 
 & KS & CM & &KS & CM & & KS & CM & & KS & CM\\
\hline
$\eta=2$ & 0.215 & 0.220 & & 0.285 & 0.310 & & 0.325 & 0.355 & & 0.455 & 0.440 \\
$\eta=3$ & 0.100 & 0.165 & & 0.175 & 0.270 & & 0.155 & 0.220 & & 0.270 & 0.295 \\
$\eta=5$ & 0.090 & 0.095 & & 0.140 & 0.150 & & 0.120 & 0.125 & & 0.190 & 0.200 \\
$\eta=10$ & 0.050 & 0.065 & & 0.091 & 0.125 & & 0.100 & 0.105 & & 0.140 & 0.190 \\
$\eta=\infty$ & 0.065 & 0.060 & & 0.105 & 0.115 & & 0.045 & 0.055 & & 0.100 & 0.100 \\ \hline 
\end{tabular}
\caption{The power of the test for verifying the validity of a heterocedastic transformation structure from Model B. The power is calculated based on 200 samples. The null hypothesis is satisfied for $\eta=\infty$.}\label{table4}
\end{center}
\end{table}

\begin{appendix}

\section{Regularity conditions}

For the asymptotic normality of the estimator $\hat\vartheta$, we need the following regularity conditions:
\begin{itemize}
\item[(a1)] $k$ is a symmetric probability density function supported on $[-1,1]$, $k$ is $d+1$ times continuously differentiable, and $k^{(j)}(\pm 1) = 0$ for $j=0,\ldots,d-1$. 

\item[(a2)] $h_j$ ($j=1,\ldots,d$) satisfies $h_j/h \rightarrow c_j$ for some $0<c_j<\infty$ and some baseline bandwidth $h$ satisfying $nh^{2p+2} \rightarrow 0$ for some $p \ge 3$, 
and $nh^{3d+\delta} \rightarrow \infty$ for some small $\delta>0$. 

\item[(a3)] The kernel $\ell$ is a symmetric, twice continuously differentiable function supported on $[-1,1]$, $\int u^s \ell(u) du = 0$ for $s=1,\ldots,q-1$ and $\int u^q \ell(u) du \neq 0$ for some $q \ge 4$. The bandwidth $g$ satisfies $ng^6 (\log n)^{-2} \rightarrow \infty$ and $ng^{2q} \rightarrow 0$.  

\item[(a4)] 
The support $R_X$ of the covariate $X$ is a compact subset of $\mathbb{R}^d$, the distribution function $F_X$ is $2d+1$-times continuously differentiable, $\inf_{x \in R_X} f_X(x) > 0$ and $\inf_{x \in R_X} \sigma(x)>0$.  Moreover, the functions $m_\vartheta(x)$, $\dot m_\vartheta(x)$, $\sigma_\vartheta(x)$ and $\dot \sigma_\vartheta(x)$ are $p+2$ times continuously differentiable with respect to the components of $x$ on $R_X \times {\cal N}(\vartheta_0)$, and all derivatives up to order $p+2$ are bounded uniformly in $(x,\vartheta) \in R_X \times {\cal N}(\vartheta_0)$, where ${\cal N}(\vartheta_0)$ is a neighborhood of $\vartheta_0$.  

\item[(a5)]
The transformation $\Lambda_{\vartheta}$ satisfies $\sup_{\vartheta \in \Theta,x \in R_X}||E[\dot{\Lambda}_{\vartheta}(Y)|X=x]|| < \infty$, \linebreak $\sup_{x \in R_X}  ||E[\dot{\Lambda}^4_{\vartheta_0}(Y) | X =x ] || < \infty$, and the density function of $(\dot{\Lambda}_{\vartheta}(Y), X)$ exists and is continuous for all $\vartheta \in \Theta$.  In addition, $\Lambda_\vartheta(y)$ is three times continuously differentiable with respect to $y$ and $\vartheta$, and there exists a $\delta>0$ such that
$$ E \Big[\sup_{\vartheta':\|\vartheta'-\vartheta\| \le \delta} \Big|\frac{\partial^{j+r}}{\partial y^j \partial \vartheta_1^{r_1} \ldots \partial \vartheta_k^{r_k}}  \Lambda_{\vartheta'}(Y) \Big| \Big] < \infty, $$
for all $\vartheta \in \Theta$ and all $0 \le j+r \le 3$, where $r=\sum_{i=1}^k r_i$.

\item[(a6)]
The error term $\eps$ has finite sixth moment and is independent of $X$.  Moreover, the distribution $F_{\eps(\vartheta)}(y)$ is three times continuously differentiable with respect to $y$ and $\vartheta$, 
$$ \sup_{y, \vartheta} \Big|\frac{\partial^{j+r}}{\partial y^j \partial \vartheta_1^{r_1} \ldots \partial \vartheta_k^{r_k}} F_{\eps(\vartheta)}(y) \Big| < \infty $$
for all $0 \le j+\sum_{i=1}^k r_i \le 2$, $\sup_y |y f_\eps^\prime(y)| < \infty$, $\sup_y |y \dot f_\eps^\prime(y)| < \infty$ and $\sup_y |y^2 f_\eps^{\prime\prime}(y)| < \infty$. 
In addition, the conditional distribution $F_{\eps(\vartheta)|X}(y|x)$ is three times continuously differentiable with respect to $y$ and $\vartheta$, 
$$ \sup_{y,x,\vartheta}  \Big|\frac{\partial^{j+r}}{\partial y^j\partial\vartheta_1^{r_1}\ldots\partial\vartheta_k^{r_k}} F_{\eps(\vartheta)|X}(y|x) \Big| <\infty $$
for all $0 \le j+\sum_{i=1}^k r_i \le 2$, $\sup_{y,x} |y f_{\eps|X}^\prime(y|x)| < \infty$, $\sup_{y,x} |y \dot f_{\eps|X}^\prime(y|x)| < \infty$ and $\sup_{y,x} |y^2 f_{\eps|X}^{\prime\prime}(y|x)| < \infty$. 

\item[(a7)] 
For all $\eta>0$, there exists $\eps(\eta)>0$ such that $\inf_{\|\vartheta-\vartheta_0\|>\eta} \|G(\vartheta)\| \ge \eps(\eta)>0$. Moreover, the matrix $\Gamma$ defined in Theorem \ref{asnotheta} is of full rank. 

\end{itemize}

\medskip

For the results of section \ref{testing}, we will need assumptions (a1), (a2) and the following conditions. Let $\|\cdot\|$ denote some vector or matrix norm, depending on the object. 

\begin{itemize}
\item[(A1)] All partial derivatives of $F_X$ up to order $2d+1$ exist on the interior of its compact support $R_X$, they are uniformly continuous and 
$\ds\inf_{x \in R_X} f_X(x) >0$. 

\item[(A2)] All partial derivatives of $m$ and $\si$ up to order $p+2$ exist on the interior of $R_X$, they are uniformly continuous and $\ds\inf_{x \in R_X} \si(x) > 0$.

\item[(A3)] $F_\ep$ is twice continuously differentiable, $\ds\sup_y|y f_\ep(y)| < \infty$, $\ds\sup_y|y^2 f'_\ep(y)| < \infty$, and $E(\ep^6) < \infty$.

\item[(A4)] $\ds \sup_{y\in\mathbb{R}}E\left[\left\|\nabla_\vartheta F_{\eps(\vartheta)|X}(y|X)|_{\vartheta=\vartheta_0}\right\|\right]<\infty$

\item[(A5)] For the parameter estimator  a linear expansion as in (\ref{lin-exp-theta}) is valid with $E[g_{\vartheta_0}(X,Y)]=0$, $E[\|g_{\vartheta_0}(X,Y)\|^2]<\infty$.

\item[(A6)] Let $F_{Y|X}(\cdot|x)$ and $f_{Y|X}(\cdot|x)$ denote the conditional distribution and density function of $Y$, given $X=x$, respectively. We assume existence of some $\eta>0$ such that
$$\sup_{\vartheta:\|\vartheta-\vartheta_0\|\leq\eta}\sup_{z\in\mathbb{R}}\int\Big(|f_{Y|X}^\prime(V_{\vartheta}(z)|u)|\|\dot V_{\vartheta}(z)\|^2+f_{Y|X}(V_{\vartheta}(z)|u)\|\ddot V_{\vartheta}(z)\|
\Big)dF_X(x)<\infty.$$
Here we use the notation $V_\vartheta=\Lambda_\vartheta^{-1}$ for the inverse of the transformation and  $\dot V_\vartheta=\nabla_\vartheta V_\vartheta$ and $\ddot V_\vartheta=(\frac{\partial^2 V_\vartheta}{\partial \vartheta_i\vartheta_j})_{i,j=1,\dots,k}$ for the gradiant and Hessian matrix, respectively. 
Further we assume that $\displaystyle \sup_{y\in\mathbb{R},x\in R_X}\Big\|y\frac{\partial(f_{Y|X}(V_{\vartheta_0}(y)|x)\dot V_{\vartheta_0}(y))}{\partial y}\Big\|<\infty$.

\item[(A7)] For some $\eta>0$, $E[\sup_{\vartheta:\|\vartheta-\vartheta_0\|\leq \eta}\|\ddot \Lambda_{\vartheta}(Y)\|]<\infty$, $E[\sup_{\vartheta:\|\vartheta-\vartheta_0\|\leq \eta}\|\dot \Lambda_{\vartheta}(Y)\|^2]<\infty$ and $E[\sup_{\vartheta:\|\vartheta-\vartheta_0\|\leq \eta}\|\ddot \Lambda_{\vartheta}(Y)\Lambda_{\vartheta}(Y)\|]<\infty$. 
Further,
\begin{eqnarray*}
&& E\Big[\sup_{\vartheta:\|\vartheta-\vartheta_0\|\leq\eta} \|\Lambda_{\vartheta}(Y)\dot\Lambda_\vartheta(Y)\|\,\Big| X=x\Big]<\infty\\
&& E\Big[\sup_{\vartheta:\|\vartheta-\vartheta_0\|\leq\eta} \|\dot\Lambda_\vartheta(Y)\|\,\Big| X=x\Big]<\infty
\end{eqnarray*}
for almost all $x\in R_X$. 

\item[(A8)] Assumption (A2) holds with $m$ replaced by $E[\frac{\partial\Lambda_{\vartheta}(Y)}{\partial\vartheta_i}|_{\vartheta=\vartheta_0}|X=\cdot]$ and $\sigma$ replaced by $E[\Lambda_{\vartheta_0}(Y)\frac{\partial\Lambda_{\vartheta}(Y)}{\partial\vartheta_i}|_{\vartheta=\vartheta_0}|X=\cdot]$, for $i=1,\dots,k$. Further, $E[\|\dot\Lambda_{\vartheta_0}(Y)\|^3]<\infty$ and $E[\|\Lambda_{\vartheta_0}(Y)\dot\Lambda_{\vartheta_0}(Y)\|^3]<\infty$. 
\end{itemize}

\section{Proof of main results}

\subsection{Proof of Theorem \ref{asnotheta}}

We will follow the different steps of the proof of Theorem 4.1 in Linton {\it et al}.\ (2008), which shows the asymptotic normality of $\hat \vartheta$ in the homoscedastic case.   However, for reasons of brevity of exposition, we will focus on the differences with respect to that proof.   The proof in Linton {\it et al}.\ (2008) consists of 11 lemmas from which the result follows.   The lemmas that need closer attention are Lemmas A.1, A.2, A.3 and A.11.  The other lemmas can be extended to the heteroscedastic case in a straightforward way. We start with the extension of Lemma A.1 to the heteroscedastic case.  This lemma develops an i.i.d.\ expansion for $\hat f_{\hat\eps(\vartheta_0)}(y) - f_{\eps(\vartheta_0)}(y)$.  For this, first note that
\begin{eqnarray}\label{wxn}
&& \hat m_{\vartheta_0}(x) = \frac{1}{nh^d}\sumi W_{x,n}\Big(\frac{x-X_i}{h}\Big) \Lambda_{\vartheta_0}(Y_i),
\end{eqnarray}
where $W_{x,n}(u) = K^*(u)/f_X(x) (1+o_P(1)) $
uniformly in $u \in [-1,1]^d$ and $x \in R_X$, and $(nh^d)^{-1} \sumi W_{x,n}((x-X_i)/h)=1$.  
The kernel $K^*(\cdot)$ is the so-called equivalent kernel 
 and is a linear combination of functions of the form $\prod_{i=1}^d k(u_i)u_i^{j_i}$ with $(j_1,\dots,j_d)\in\mathbb{N}_0^d$, $0\leq\sum_{i=1}^d j_i\leq p$.  This can be deduced from representation (3.25) in combination with (3.30), (3.9) and (3.19) in Gu, Li and Yang (2014); see also Masry (1996a, 1996b) and 
Fan and Gijbels (1996), p.\ 63--64, for the case $d=1$.  In a similar way we can also write
\begin{eqnarray}\label{wxn2}
\hat \sigma_{\vartheta_0}(x) - \sigma_{\vartheta_0}(x) &=& \frac{1}{2\sigma_{\vartheta_0}(x)} \frac{1}{nh^d} \sumi W_{x,n} \Big(\frac{x-X_i}{h}\Big) \Big[(\Lambda_{\vartheta_0}(Y_i)-m_{\vartheta_0}(x))^2 - \sigma_{\vartheta_0}^2(x)\Big] \nonumber \\
&& + o_P(n^{-1/2}).
\end{eqnarray}
It follows that we can write
\begin{eqnarray*}
&& \hat f_{\hat\eps(\vartheta_0)}(y) - f_{\eps(\vartheta_0)}(y) \\
&& = \frac{1}{ng} \sumi \ell_g^\prime(\eps_i-y) (\hat \eps_i(\vartheta_0)-\eps_i) + \frac{1}{n} \sumi \ell_g(\eps_i-y) - f_\eps(y) + o_P(n^{-1/2}) \\
&& = -\frac{1}{ng} \sumi \frac{\ell_g^\prime(\eps_i-y)}{\sigma(X_i)} \Big\{[\hat m_{\vartheta_0}(X_i) - m_{\vartheta_0}(X_i)] + \eps_i[\hat \sigma_{\vartheta_0}(X_i) - \sigma_{\vartheta_0}(X_i)] \Big\} \\
&& \hspace*{.5cm} + \frac{1}{n} \sumi \ell_g(\eps_i-y) - f_\eps(y) + o_P(n^{-1/2}) \\
&& = (T_1+T_2)(y) + o_P(n^{-1/2}) \hspace*{1cm} \mbox{(say)}.
\end{eqnarray*}
Using decompositions (\ref{wxn}) and (\ref{wxn2}), we have that
\begin{eqnarray*}
T_1(y) &=& -\frac{1}{ng} \sumi \frac{\ell_g^\prime(\eps_i-y)}{\sigma(X_i)} \frac{1}{nh^d} \sumj W_{X_i,n} \Big(\frac{X_i-X_j}{h}\Big) (\Lambda_{\vartheta_0}(Y_j)-m_{\vartheta_0}(X_i)) \\
&& -\frac{1}{ng} \sumi \frac{\ell_g^\prime(\eps_i-y)}{\sigma^2(X_i)} \frac{\eps_i}{2nh^d} \sumj W_{X_i,n} \Big(\frac{X_i-X_j}{h}\Big) \Big((\Lambda_{\vartheta_0}(Y_j)-m_{\vartheta_0}(X_i))^2 - \sigma_{\vartheta_0}^2(X_i) \Big) \\
&& + o_P(n^{-1/2}) \\
&=& \frac{1}{n^2} \sumi \sumj A_{nij} (\eps_j + \frac{\eps_i}{2}(\eps_j^2-1)) + o_P(n^{-1/2}),
\end{eqnarray*}
where $A_{nij} = -(gh^d)^{-1} \ell_g^\prime(\eps_i-y) W_{X_i,n}((X_i-X_j)/h)$.  Using similar arguments as in Linton {\it et al}.\ (2008) and Colling and Van Keilegom (2014), the last expression can be written as
$$ f_{\eps(\vartheta_0)}^\prime(y) \frac{1}{n} \sumi \eps_i + \Big(yf_{\eps(\vartheta_0)}^\prime(y) + f_{\eps(\vartheta_0)}(y)\Big) \frac{1}{2n} \sumi (\eps_i^2-1) + o_P(n^{-1/2}). $$

In a similar way i.i.d.\ expansions for $\dot{\hat f}_{\hat\eps(\vartheta_0)}(y) - \dot{f}_{\eps(\vartheta_0)}(y)$ and $\hat f_{\hat\eps(\vartheta_0)}^\prime(y) - f_{\eps(\vartheta_0)}^\prime(y)$ can be obtained, which then extend Lemmas A.2 and A.3 in Linton {\it et al}.\ (2008) to the heteroscedastic case.  

These three i.i.d.\ expansions all come together when we develop the i.i.d.\ expansion for $\hat\vartheta-\vartheta_0$.   For the homoscedastic case this is done in Lemma A.11 in Linton {\it et al}.\ (2008), and it is shown there that all terms that come from the estimation of $m$, $\dot m$, $f_\eps$, $f_\eps^\prime$ and $\dot f_\eps$ cancel and one therefore obtains the same expansion as in the case where all these functions would be known.   In our  heteroscedastic model a similar development can be done by using the above expansions for $\hat f_{\hat\eps(\vartheta_0)}$, $\dot{\hat f}_{\hat\eps(\vartheta_0)}$ and $\hat f_{\hat\eps(\vartheta_0)}^\prime$.  We find in a similar way as in the homoscedastic case that all these expansions cancel out, and hence we get asymptotically the same i.i.d.\ expansion as in the case where these functions would be known.   This shows the first part of Theorem \ref{asnotheta}.  The second part follows immediately from the central limit theorem, together with the fact that $E[g_{\vartheta_0}(X,Y)] = G(\vartheta_0)=0$. 
\hfill $\Box$

\subsection{Proof of Theorem \ref{theo1}}

Let $\hat F_{X,\eps}$ denote the joint empirical distribution function of $(X_i,\eps_i)$, $i=1,\dots,n$, under $H_0$. 
Let further 
\begin{eqnarray*}
R_n(x,y) &=& E[I\{X\leq x\}I\{\Lambda_{\hat\vartheta}(Y)\leq y\hat\sigma(X)+\hat m(X)\} \mid \Y_n]-E[I\{X\leq x\}I\{\eps\leq y\}],
\end{eqnarray*}
where $\Y_n=\{(X_i,Y_i)\mid i=1,\dots,n\}$. Then we have the following Lemma.

\blem\label{lem1} Under the assumptions of Theorem \ref{theo1},
$$\hat F_{X,\hat\eps}(x,y)=\hat F_{X,\eps}(x,y)+R_n(x,y)+o_P(\frac{1}{\sqrt{n}})$$
uniformly with respect to $x\in R_X,y\in\mathbb{R}$.
\elem

\noindent
{\bf Proof of Lemma \ref{lem1}}
With the definition in Proposition \ref{prop1} we have
\begin{eqnarray*}
\sqrt{n}(\hat F_{X,\hat\eps}(x,y)-\hat F_{X,\eps}(x,y)-R_n(x,y)) &=& G_n(x,\hat\vartheta,(\hat m-m)/\sigma,\hat\sigma/\sigma,y), 
\end{eqnarray*}
where the empirical process
\begin{eqnarray*}
G_n(x,\vartheta,g_1,g_2,y) &=& \frac{1}{\sqrt{n}}\sum_{i=1}^n \Big(I\{X_i\leq x\}\varphi_{\vartheta,g_1,g_2,y}(X_i,Y_i)-E[I\{X\leq x\}\varphi_{\vartheta,g_1,g_2,y}(X,Y)]\Big)
\end{eqnarray*}
(indexed in $x\in R_X,\vartheta\in\Theta, g_1\in \G_1, g_2\in \G_2,y\in\mathbb{R}$)  converges weakly to a Gaussian process. This follows from Proposition \ref{prop1}, the Donsker property of $\{I\{X\leq x\}\mid x\in R_X\}$ and because products of uniformly bounded Donsker classes are Donsker (see Example 2.10.8 in van der Vaart \& Wellner, 1996, p.\ 192). 
Thus $G_n$ is asymptotically stochastically equicontinuous with respect to 
\begin{eqnarray*}
&&\rho\Big((x,\vartheta,g_1,g_2,y),(x',\vartheta',g_1',g_2',y'))\\
&=&\Big(\Var\Big(I\{X\leq x\}\varphi_{\vartheta,g_1,g_2,y}(X,Y)-I\{X\leq x'\}\varphi_{\vartheta',g_1',g_2',y'}(X,Y)\Big)\Big)^{1/2}
\end{eqnarray*}
(see van der Vaart, 1998, p.\ 262/263). 
We have
\begin{eqnarray*}
&&\rho\Big((x,\hat\vartheta,(\hat m-m)/\sigma,\hat\sigma/\sigma,y),(x,\vartheta_0,0,1,y)\Big)= o_P(\delta_n)
\end{eqnarray*}
where $\delta_n\searrow 0$ by Proposition \ref{prop3}. Thus and because $\varphi_{\vartheta_0,0,1,y}\equiv 0$ it follows that
\begin{eqnarray*}
&&P\Big(\sup_{x,y}|\sqrt{n}(\hat F_{X,\hat\eps}(x,y)-\hat F_{X,\eps}(x,y)-R_n(x,y))| >\eta\Big)\\
&\leq& P\Big( \sup_{\rho((x,\vartheta,g_1,g_2,y),(x',\vartheta',g_1',g_2',y'))\leq \delta_n}|G_n(x,\vartheta,g_1,g_2,y)-G_n(x',\vartheta',g_1',g_2',y')|>\eta\Big)
\end{eqnarray*}
which converges to zero for $n\to\infty$, for all $\eta>0$. From this the assertion of Lemma \ref{lem1} follows. 
\hfill $\Box$

\bigskip

To finish the proof of Theorem \ref{theo1} we decompose $R_n=A_n+B_n+C_n$, where
\begin{eqnarray*}
A_n(x,y)&=& \quad E[I\{X\leq x\}I\{\Lambda_{\hat\vartheta}(Y)\leq y\hat\sigma(X)+\hat m(X)\} \mid \Y_n]\\
&&{}-E[I\{X\leq x\}I\{\Lambda_{\vartheta_0}(Y)\leq y\hat\sigma(X)+\hat m(X)\} \mid \Y_n]\\
B_n(x,y) &=&\quad E[I\{X\leq x\}I\{\Lambda_{\vartheta_0}(Y)\leq y\hat\sigma_{\hat\vartheta}(X)+\hat m_{\hat\vartheta}(X)\} \mid \Y_n]\\
&&{}-E[I\{X\leq x\}I\{\Lambda_{\vartheta_0}(Y)\leq y\hat\sigma_{\vartheta_0}(X)+\hat m_{\vartheta_0}(X)\} \mid \Y_n]\\
C_n(x,y) &=&\quad E[I\{X\leq x\}I\{\Lambda_{\vartheta_0}(Y)\leq y\hat\sigma_{\vartheta_0}(X)+\hat m_{\vartheta_0}(X)\} \mid \Y_n]\\
&&{}-E[I\{X\leq x\}I\{\Lambda_{\vartheta_0}(Y)\leq y\sigma_{\vartheta_0}(X)+m_{\vartheta_0}(X)\}].
\end{eqnarray*}
For the ease of notation in the following let the parameter $\vartheta$ be one-dimensional. 
We use the same notations as in assumption (A6). Then we have
\begin{eqnarray*}
A_n(x,y)&=& \int \Big( F_{Y|X}(V_{\hat\vartheta}(y\hat\sigma(u)+\hat m(u))|u)
-F_{Y|X}(V_{\vartheta_0}(y\hat\sigma(u)+\hat m(u))|u)\Big)I\{u\leq x\}\,dF_X(u).
\end{eqnarray*}
For the moment fix $u$ and $z=y\hat\sigma(u)+\hat m(u)$ and consider a second order Taylor expansion of the map $\vartheta\mapsto\psi(\vartheta)= F_{Y|X}(V_{\vartheta}(z)|u)$, i.\,e.\
\begin{eqnarray*}
\psi(\hat\vartheta)-\psi(\vartheta_0)&=&f_{Y|X}(V_{\vartheta_0}(z)|u)\dot V_{\vartheta_0}(z)(\hat\vartheta-\vartheta_0)\\
&&{}+\frac{1}{2}\Big(f_{Y|X}^\prime(V_{\vartheta^*}(z)|u)(\dot V_{\vartheta^*}(z))^2+f_{Y|X}(V_{\vartheta^*}(z)|u)\ddot V_{\vartheta^*}(z)
\Big)(\hat\vartheta-\vartheta_0)^2.
\end{eqnarray*}
The value $\vartheta^*$ may depend on $u$ and $z$, but lies between $\hat\vartheta$ and $\vartheta_0$. Because for each $\eta>0$, $|\hat\vartheta-\vartheta_0|\leq\eta$ with probability converging to one, for the proof we may assume $|\vartheta^*-\vartheta_0|\leq \eta$ with $\eta$ from assumption (A6).
A Taylor expansion of $\psi$ motivates the definition of 
\begin{eqnarray*}
\tilde A_n(x,y)&=& \int f_{Y|X}(V_{\vartheta_0}(y\hat\sigma(u)+\hat m(u))|u)\dot V_{\vartheta_0}(y\hat\sigma(u)+\hat m(u))I\{u\leq x\}\,dF_X(u)(\hat\vartheta-\vartheta_0)
\end{eqnarray*}
and yields that 
\begin{eqnarray*}
&&\sup_{x,y}|A_n(x,y)-\tilde A_n(x,y)|\\
&\leq& (\hat\vartheta-\vartheta_0)^2\frac{1}{2}\sup_{\vartheta:|\vartheta-\vartheta_0|\leq\eta}\sup_{z\in\mathbb{R}}\int\Big(|(f_{Y|X}^\prime(V_{\vartheta}(z)|u)|(\dot V_{\vartheta}(z))^2+f_{Y|X}(V_{\vartheta}(z)|u)|\ddot V_{\vartheta}(z)| \Big) dF_X(x)\\
&=& o_P(\frac{1}{\sqrt{n}})
\end{eqnarray*}
by assumption (A6).
Denote by $\bar A_n$ the same term as $\tilde A_n$, but with the estimators $\hat\sigma$ and $\hat m$ replaced by the true functions $\sigma$ and $m$, respectively.
Note that from the proof of Proposition \ref{prop2} uniform convergence of $|\hat\sigma-\sigma|$ and $|\hat m-m|$ to zero in probability follows and thus by the mean value theorem, the last part of assumption (A6), and $\hat\vartheta-\vartheta_0=O_P(n^{-1/2})$ we obtain $\sup_{x,y}|\tilde A_n(x,y)-\bar A_n(x,y)|=o_P(n^{-1/2})$. 
Altogether for $A_n$ we have uniformly with respect to $x\in R_X$, $y\in\mathbb{R}$,
\begin{eqnarray*}
A_n(x,y)&=& \int f_{Y|X}(V_{\vartheta_0}(y\sigma(u)+ m(u))|u) \dot V_{\vartheta_0}(y\sigma(u)+ m(u))I\{u\leq x\}\,dF_X(u)(\hat\vartheta-\vartheta_0)\\
&&{}+o_P(\frac{1}{\sqrt{n}}).
\end{eqnarray*} 
 
For $C_n$ we obtain the following expansion uniformly with respect to $x,y$,
\begin{eqnarray*}\nonumber
C_n(x,y) &=&\quad E \Big[I\{X\leq x\}I\Big\{\eps\leq y\frac{\hat\sigma_{\vartheta_0}(X)}{\sigma(X)}+\frac{\hat m_{\vartheta_0}(X)-m(X)}{\sigma(X)}\Big\} \mid \Y_n\Big]\\
\nonumber
&&{}-E[I\{X\leq x\}I\{\eps\leq y\}]\\
\nonumber
&=& \int \Big(F_\eps\Big(y\frac{\hat\sigma_{\vartheta_0}(u)}{\sigma(u)}+\frac{\hat m_{\vartheta_0}(u)-m(u)}{\sigma(u)}\Big)-F_\eps(y)\Big) I\{u \le x\} \,d F_X(u)\\
\nonumber
&=& f_\eps(y)\Big( y\int \frac{\hat\sigma_{\vartheta_0}(u)-\sigma(u)}{\sigma(u)}I\{u\leq x\}\,dF_X(u)\\
\nonumber
&&{}+\int \frac{\hat m_{\vartheta_0}(u)-m(u)}{\sigma(u)}I\{u\leq x\}\,dF_X(u)\Big) +o_P(\frac{1}{\sqrt{n}})\\
\label{Bn-ent}
&=& f_\eps(y)\frac 1n\sum_{i=1}^n (\eps_i+\frac{y}{2}(\eps_i^2-1))\int \frac{1}{h}K^*\Big(\frac{u-X_i}{h}\Big)I\{u\leq x\}\,du
+o_P(\frac{1}{\sqrt{n}}).
\end{eqnarray*}
The second but last equality follows by Taylor's expansion, assumption (A3) and the fact that $\int (\hat m_{\vartheta_0}-m)^2/\sigma^2\,dF_X=o_P(n^{-1/2})$, $\int (\hat \sigma_{\vartheta_0}-\sigma)^2/\sigma^2\,dF_X=o_P(n^{-1/2})$, see the proof of Theorem 2.1 in Neumeyer and Van Keilegom (2010). The last equality follows from (\ref{wxn}) and (\ref{wxn2}), a combination of the proof of Lemma A.2 in Neumeyer and Van Keilegom (2010), and the proof of Proposition 2 (p.\ 537) in Neumeyer and Van Keilegom (2009). 

Now let either $Z_i=\eps_i$ or $Z_i=\eps_i^2-1$. Then exactly as in the last part of the proof of Lemma B.1 in the supporting information to Birke and Neumeyer (2013) we have
\begin{eqnarray*}\label{mel}
\sup_{x\in R_X}\Big|\frac 1n\sum_{i=1}^n Z_i \Big(\int \frac{1}{h^d}K^*\Big(\frac{u-X_i}{h}\Big)I\{u\leq x\}\, du-I\{X_i\leq x\}\Big)\Big|=o_P(\frac{1}{\sqrt{n}}).
\end{eqnarray*}
Altogether for $C_n$ we have uniformly with respect to $x\in R_X$, $y\in\mathbb{R}$,
\begin{eqnarray*}
 C_n(x,y)&=& f_\eps(y)\frac 1n\sum_{i=1}^n (\eps_i+\frac{y}{2}(\eps_i^2-1))I\{X_i\leq x\}+o_P(\frac{1}{\sqrt{n}}).
\end{eqnarray*}

With $B_n$ we proceed similarly to obtain
\begin{eqnarray*}
B_n(x,y) 
&=& f_\eps(y)\Big( y\int \frac{\hat\sigma_{\hat\vartheta}(u)-\hat\sigma_{\vartheta_0}(u)}{\sigma(u)}I\{u\leq x\}\,dF_X(u)\\
\nonumber
&&{}+\int \frac{\hat m_{\hat\vartheta}(u)-\hat m_{\vartheta_0}(u)}{\sigma(u)}I\{u\leq x\}\,dF_X(u)\Big) +o_P(\frac{1}{\sqrt{n}})
\end{eqnarray*}
by assumption (A3) and the fact that $\sup_x|\hat m_{\hat\vartheta}(x)-\hat m_{\vartheta_0}(x)|=O_P(n^{-1/2})$, $\sup_x|\hat \sigma_{\hat\vartheta}(x)-\hat \sigma_{\vartheta_0}(x)|=O_P(n^{-1/2})$ (see the proof of Proposition \ref{prop2}). 
Now note that 
\begin{eqnarray}\label{hatm-entw}
\hat m_{\hat\vartheta}(u)-\hat m_{\vartheta_0}(u) &=& \frac{1}{nh^d}\sumi W_{u,n}\Big(\frac{u-X_i}{h}\Big) (\Lambda_{\hat\vartheta}(Y_i)-\Lambda_{\vartheta_0}(Y_i))\\
&=& \frac{1}{nh^d}\sumi W_{u,n}\Big(\frac{u-X_i}{h}\Big) \dot\Lambda_{\vartheta_0}(Y_i)(\hat\vartheta-\vartheta_0)+r_n(u),\nonumber
\end{eqnarray}
where 
\begin{eqnarray*}
&& \int \frac{r_n(u)}{\sigma(u)}I\{u\leq x\}\,dF_X(u) \\
&& \leq \frac12 (\hat\vartheta-\vartheta_0)^2 \int \frac{1}{nh^d}\sumi \Big|W_{u,n}\Big(\frac{u-X_i}{h}\Big)\Big| \sup_{\vartheta:|\vartheta-\vartheta_0|\leq \eta}|\ddot \Lambda_{\vartheta}(Y_i)| \frac{I\{u \le x\}}{\sigma(u)} \,dF_X(u)\\
&& = o_P(n^{-1/2})
\end{eqnarray*}
by assumptions (A5) and (A7). Proceeding similarly to the expansion of $C_n$ we thus obtain
\begin{eqnarray*}
&&\int \frac{\hat m_{\hat\vartheta}(u)-\hat m_{\vartheta_0}(u)}{\sigma(u)}I\{u\leq x\}\,dF_X(u)\\
&=& (\hat\vartheta-\vartheta_0) \frac 1n\sum_{i=1}^n \dot\Lambda_{\vartheta_0}(Y_i)\int \frac{1}{h^d}K^*\Big(\frac{u-X_i}{h}\Big)\frac{I\{u\leq x\}}{\sigma(u)}\,dx+o_P(\frac{1}{\sqrt{n}})\\
&=& (\hat\vartheta-\vartheta_0) E\Big[\dot\Lambda_{\vartheta_0}(Y)\frac{I\{X\leq x\}}{\sigma(X)}\Big]+o_P(\frac{1}{\sqrt{n}}).
\end{eqnarray*}
Similarly for the variance we have $\hat\sigma_{\hat\vartheta}-\hat\sigma_{\vartheta_0}=(\hat\sigma^2_{\hat\vartheta}-\hat\sigma^2_{\vartheta_0})/(\hat\sigma_{\hat\vartheta}+\hat\sigma_{\vartheta_0})$
which yields (compare to (\ref{hatm-entw}))
\begin{eqnarray*}
&&\int \frac{\hat \sigma_{\hat\vartheta}(u)-\hat \sigma_{\vartheta_0}(u)}{\sigma(u)}I\{u\leq x\}\,dF_X(u)\\
&=&\frac{1}{2}\int\frac{1}{\sigma^2(u)} \frac{1}{nh^d}\sum_{i=1}^n W_{u,n}\Big(\frac{u-X_i}{h}\Big) ((\Lambda_{\hat\vartheta}(Y_i))^2-(\Lambda_{\vartheta_0}(Y_i))^2)I\{u\leq x\}\,dF_X(u)
\\
&&{}+\frac{1}{2}\int\frac{1}{\sigma^2(u)} (\hat m_{\vartheta_0}(u)-\hat m_{\hat\vartheta}(u))(\hat m_{\vartheta_0}(u)+\hat m_{\hat\vartheta}(u))I\{u\leq x\}\,dF_X(u)
+o_P(\frac{1}{\sqrt{n}})\\
&=&  (\hat\vartheta-\vartheta_0)\Bigg( \frac {1}{2n}\sum_{i=1}^n \frac{\partial (\Lambda_{\vartheta}(Y_i))^2}{\partial\vartheta}\Big|_{\vartheta=\vartheta_0}
\int \frac{1}{h^d}K^*\Big(\frac{u-X_i}{h}\Big)\frac{I\{u\leq x\}}{\sigma^2(u)}\,du\\
&&{}-  \frac{1}{2n} \sum_{i=1}^n\dot\Lambda_{\vartheta_0}(Y_i) 
\int \frac{1}{h^d}K^*\Big(\frac{u-X_i}{h}\Big)\frac{I\{u\leq x\}}{\sigma^2(u)}2m(u)\,du \Bigg)
+o_P(\frac{1}{\sqrt{n}})\\
&=& (\hat\vartheta-\vartheta_0) \frac{1}{n}\sum_{i=1}^n \Big( \dot\Lambda_{\vartheta_0}(Y_i)\Lambda_{\vartheta_0}(Y_i)- \dot\Lambda_{\vartheta_0}(Y_i)m(X_i)\Big)\frac{I\{X_i\leq x\}}{\sigma^2(X_i)}+o_P(\frac{1}{\sqrt{n}})\\
&=& (\hat\vartheta-\vartheta_0) E\Big[\Big(\dot\Lambda_{\vartheta_0}(Y)\Lambda_{\vartheta_0}(Y)-\dot\Lambda_{\vartheta_0}(Y)m(X)\Big)\frac{I\{X\leq x\}}{\sigma^2(X)}\Big]+o_P(\frac{1}{\sqrt{n}}).
\end{eqnarray*}
Those expansions yield  uniformly with respect to $x$ and $y$, 
\begin{eqnarray*}
B_n(x,y) 
&=& (\hat\vartheta-\vartheta_0) f_\eps(y) E\Big[\dot\Lambda_{\vartheta_0}(Y)\Big(\sigma(X)+y\Lambda_{\vartheta_0}(Y)-ym(X)\Big)\frac{I\{X\leq x\}}{\sigma^2(X)}\Big]\\
&&{}
+o_P(\frac{1}{\sqrt{n}}).
\end{eqnarray*}
The expansions derived for $A_n$, $B_n$ and $C_n$ now yield
\begin{eqnarray}\label{entw}
R_n(x,y) 
&=& (\hat\vartheta-\vartheta_0)H_{\vartheta_0}(x,y)+ f_\eps(y)\frac 1n\sum_{i=1}^n (\eps_i+\frac{y}{2}(\eps_i^2-1))I\{X_i\leq x\}\\
&&{}+o_P(\frac{1}{\sqrt{n}})\nonumber
\end{eqnarray}
with 
\begin{eqnarray*}
H_{\vartheta_0}(x,y) 
&=&  f_\eps(y) E\Big[\dot\Lambda_{\vartheta_0}(Y)\Big(\sigma(X)+y\Lambda_{\vartheta_0}(Y)-ym(X)\Big)\frac{I\{X\leq x\}}{\sigma^2(X)}\Big]\\
&&{}+\int f_{Y|X}(V_{\vartheta_0}(y\sigma(u)+ m(u))|u) \dot V_{\vartheta_0}(y\sigma(u)+ m(u))I\{u\leq x\}\,dF_X(u)\\
&=& E\Big[\frac{\partial}{\partial\vartheta}F_{\eps(\vartheta)|X}(y|X)\Big|_{\vartheta=\vartheta_0}I\{X\leq x\}\Big].
\end{eqnarray*}
The last equality follows by some tedious but straightforward calculations.
Now the assertion of Theorem \ref{theo1} follows by Lemma \ref{lem1}, (\ref{entw}) and  assumption (A5). \hfill $\Box$

\subsection{Proof of Corollary \ref{cor1}}
From expansion (\ref{sncor}) we have
$$S_n(x,y)=G_n\Big(x,y,f_\eps(y),yf_\eps(y),h_{\vartheta_0}(x,y)\Big)+o_P(1)$$
uniformly, where
$$h_{\vartheta_0}(x,y)=E\Big[\nabla_{\vartheta}F_{\eps(\vartheta)|X}(y|X)\Big|_{\vartheta=\vartheta_0}\Big(I\{X\leq x\}-F_X(x)\Big)\Big]$$
and where the process
\begin{eqnarray*}
&&G_n(x,y,z_1,z_2,z_3)\\
&=&\frac{1}{\sqrt{n}}\sum_{i=1}^n \Big(\Big(I\{X_i\leq x\}-F_X(x)\Big)\Big(I\{\eps_i\leq y\}-F_\eps(y)+z_1\eps_i+\frac{z_2}{2}(\eps_i^2-1)\Big)\\
&&{}+z_3 g_{\vartheta_0}(X_i,Y_i)\Big),
\end{eqnarray*}
is indexed in $\F=\{(x,y,z_1,z_2,z_3)\mid x\in R_X,y\in\mathbb{R},z_1,z_2,z_3\in[-K,K]\}$ for some $K$ such that
$\sup_y f_\eps(y)\leq K,\sup_y|yf_\eps(y)|\leq K,\sup_{x,y}|h_{\vartheta_0}(x,y)|\leq K$ (see assumptions (A3) and (A4)). 
Weak convergence of $G_n$ follows similarly to the proof of Theorem 2 in Neumeyer and Van Keilegom (2009, p.\ 538). The key argument is that for the bracketing number $N_{[]}(\eta,\F,L_2(P))$ an order $O(\eta^{-7})$ can be derived from the $L_2(P)$-norm
\begin{eqnarray*}
&&\Big(E\Big[\Big(\Big(I\{X_i\leq x\}-F_X(x)\Big)\Big(I\{\eps_i\leq y\}-F_\eps(y)+z_1\eps_i+\frac{z_2}{2}(\eps_i^2-1)\Big)+z_3 g_{\vartheta_0}(X_i,Y_i)\\
&&
-\Big(I\{X_i\leq x'\}-F_X(x')\Big)\Big(I\{\eps_i\leq y'\}-F_\eps(y')+z_1'\eps_i+\frac{z_2'}{2}(\eps_i^2-1)\Big)-z_3' g_{\vartheta_0}(X_i,Y_i)\Big)^2\Big]\Big)^{1/2}\\
&& \leq C\Big(
|F_X(x)-F_X(x')|(1+K^2(1+\Var(\eps^2)))+|F_\eps(y)-F_\eps(y')|+(z_1-z_1')^2\\
&& \hspace*{.5cm} +(z_2-z_2')^2\Var(\eps^2)+(z_3-z_3')^2E[g_{\vartheta_0}^2(X,Y)]
\Big)^{1/2}
\end{eqnarray*}
for some constant $C$.
Weak convergence of $S_n$ follows by consideration of the subclass of $\F$ defined by $z_1=f_\eps(y),z_2=yf_\eps(y),z_3=h_{\vartheta_0}(x,y)$. 
\hfill $\Box$

\section{Auxiliary results}

Let for $k = (k_1,\ldots,k_d)\in\mathbb{N}_0^d$, $k.=\sum_{j=1}^d k_j$, $ D^k = \partial^{k.}/\partial x_1^{k_1} \ldots \partial x_d^{k_d}$, and
 $$ \|f\|_{d+\alpha} = \max_{k. \leq d} \sup_{x \in R_X} |D^kf(x)| + \max_{k.=d} \sup_{x,x' \in R_X} \frac{|D^kf(x)-D^kf(x')|}{\|x-x'\|^{\alpha}}, $$
where $\|\cdot\|$ is the Euclidean norm on $\mathbb{R}^d$.
Let further $\G_1=C_1^{d+\alpha}(R_X)$ be the class of $d$ times differentiable functions $f$ defined on $R_X$ such that 
$ \|f\|_{d+\alpha} \le 1,$
and $\G_2=\tilde C_2^{d+\alpha}(R_X)$ be the class of $d$ times differentiable functions $f$ defined on $R_X$ such that $\|f\|_{d+\alpha} \le 2$ and $\inf_{x \in R_X} f(x) \ge 1/2$.  

\bpro\label{prop1}
Let $\F=\{\varphi_{\vartheta, g_1,g_2,y}\mid \vartheta\in\Theta, g_1\in\G_1,g_2\in \G_2, y\in\mathbb{R}\}$, where
$$\varphi_{\vartheta, g_1,g_2,y}(X,Y)= I\Big\{\frac{\Lambda_\vartheta(Y)-m(X)}{\sigma(X)}\leq y g_2(X)+g_1(X)\Big\} -I\Big\{\frac{\Lambda_{\vartheta_0}(Y)-m(X)}{\sigma(X)}\leq y\Big\}$$
is a function from $R_X\times \mathbb{R}$ to $\mathbb{R}$ and $\G_1$, $\G_2$ are defined above. 
Then $\F$ is Donsker.
\epro

\noindent
{\bf Proof of Proposition \ref{prop1}}
In Lemma 1 in Heuchenne {\it et al.} (2014) the special case of  univariate $X$  and $\sigma\equiv 1$ (i.\,e.\ homoscedasticity) is considered. For the subclass of $\F$ obtained by setting $g_2\equiv 1$ the assertion is proved. On the other hand  Lemma A.3 in Neumeyer and Van Keilegom (2010) shows the assertion for the function class defined analogously to $\F$, but replacing $\Lambda_{\vartheta}$ by the identity (for multivariate $X$). A detailed proof combines
 the arguments of both proofs but is omitted for the sake of brevity. 
\hfill $\Box$

\bpro\label{prop2}
For the estimators $\hat m$ and $\hat \sigma$ defined in section \ref{estimation} and the function classes $\G_1$, $\G_2$ defined above we have under the assumptions of Theorem \ref{theo1} that $P((\hat m-m)/\sigma\in\G_1)\to 1$ and $P(\hat \sigma/\sigma\in\G_2)\to 1$ for $n\to\infty.$
\epro

\noindent
{\bf Proof of Proposition \ref{prop2}} 
Note that the assertion follows from  $\|\hat m-m\|_{d+\alpha}=o_P(1)$ and  $\|\hat \sigma-\sigma\|_{d+\alpha}=o_P(1)$. 
Further note that 
$$\hat m-m=(\hat m_{\vartheta_0}-m)+(\hat m_{\hat\vartheta}-\hat m_{\vartheta_0}),\quad
\hat\sigma-\sigma=(\hat \sigma_{\vartheta_0}-\sigma)+(\hat \sigma_{\hat\vartheta}-\hat \sigma_{\vartheta_0})$$ and that  $\|\hat m_{\vartheta_0}-m\|_{d+\alpha}=o_P(1)$, $\|\hat \sigma_{\vartheta_0}-\sigma\|_{d+\alpha}=o_P(1)$ was shown in Lemma A.1 in Neumeyer and Van Keilegom (2010) under assumptions (a1), (a2), (A1)--(A3). We will apply Taylor expansions for the remainder terms. To this end due to $\hat\vartheta=\vartheta_0+o_P(1)$ (see assumption (A5)) we may assume that $\|\hat\vartheta-\vartheta_0\|\leq\eta$ for $\eta$ from assumption (A7).
Denote by $\widehat{\tilde m}_{\vartheta_0}$ a local polynomial estimator defined analogously to $\hat m_{\vartheta_0}$, but based on the sample $(X_i,\dot\Lambda_{\vartheta_0}(Y_i))$, $i=1,\dots,n$. Let, by slight abuse of notation, 
$$d^kV_{x,n}(z)=\frac{\partial^{k.}(W_{x,n}(\frac{x-z}{h}))}{\partial x_1^{k_1}\dots\partial x_d^{k_d}}$$
for $k=(k_1,\dots,k_d)\in\mathbb{N}_0^d$, with $W_{x,n}$ from (\ref{wxn}). 
Then we obtain from (\ref{hatm-entw}) that
\begin{eqnarray}\nonumber
&&\|\hat m_{\hat\vartheta}-\hat m_{\vartheta_0}\|_{d+\alpha}\\
&\leq &  \|\hat\vartheta-\vartheta_0\|\|\widehat{\tilde m}_{\vartheta_0}\|_{d+\alpha}\label{masry1}\\
&&{}+\frac12 \|\hat\vartheta-\vartheta_0\|^2 \max_{k. \leq d} \frac{1}{nh^d}\sum_{i=1}^n \sup_{x \in R_X}|d^kV_{x,n}(X_i)|\sup_{\|\vartheta-\vartheta_0\|\leq \eta}\|\ddot\Lambda_{\vartheta}(Y_i)\|\label{masry2}\\
&&{}+\frac12 \|\hat\vartheta-\vartheta_0\|^2\max_{k.=d}\frac{1}{nh^d} \sum_{i=1}^n \sup_{x,x' \in R_X}\frac{|d^kV_{x,n}(X_i)-d^kV_{x',n}(X_i)|}{\|x-x'\|^{\alpha}}\sup_{\|\vartheta-\vartheta_0\|\leq \eta}\|\ddot\Lambda_{\vartheta}(Y_i)\|.\qquad\quad\label{masry3}
\end{eqnarray}
Under assumptions (a1), (a2), (A1) and (A8) we have that $\|\widehat{\tilde m}_{\vartheta_0}\|_{d+\alpha}$ converges to $\|\tilde m_{\vartheta_0}\|_{d+\alpha}$ in probability, where $\tilde m_{\vartheta_0}(\cdot)=E[\dot\Lambda_{\vartheta_0}(Y)|X=\cdot]$. Thus (\ref{masry1}) is negligible since $\|\hat\vartheta-\vartheta_0\|=O_P(n^{-1/2})$. 
Under assumptions (a1) and (a2),  from the representations of the multivariate local polynomial estimator in Masry (1996a, 1996b) one can deduce that $h^{d}\sup_{x,z}|d^kV_{x,n}(z)|$ is bounded (for $k.\leq d$).  Thus applying the law of large numbers to $\sup_{\|\vartheta-\vartheta_0\|\leq \eta}\|\ddot\Lambda_{\vartheta}(Y_i)\|$ (compare to assumption (A7)) for (\ref{masry2}) we obtain the order $O_P(\|\hat\vartheta-\vartheta_0\|^2h^{-2d})=o_P(1)$ by assumption (a2). Further, by considering the cases $\|x-x'\|\geq h$ and $\|x-x'\|< h$ one obtains
\begin{eqnarray*}
\sup_{x,x' \in R_X}\frac{|d^kV_{x,n}(X_i)-d^kV_{x',n}(X_i)|}{\|x-x'\|^{\alpha}}
&\leq&
2\sup_{x,z}|d^k V_{x,n}(z)|\frac{1}{h^\alpha}
+\sum_{j=1}^d\sup_{x,z}\Big|\frac{\partial d^kV_{x,n}(z)}{\partial x_j}\Big| h^{1-\alpha}.
\end{eqnarray*}
All partial derivatives of order one of $h^{d+1}d^kV_{x,n}(z)$ in $x$-direction are bounded in $x,z$. Thus for (\ref{masry3}) one obtains the rate
$O_P(\|\hat\vartheta-\vartheta_0\|^2(h^{-(2d+\alpha)}+h^{-(2d+1-(1-\alpha)}))=o_P(1)$ by assumption (a2). 
Similar arguments hold for $\hat\sigma_{\hat\vartheta}-\hat\sigma_{\vartheta_0}$.
 \hfill $\Box$

\bpro\label{prop3}
With the definitions in Proposition \ref{prop1} we have under the assumptions of Theorem \ref{theo1} that $E[(\varphi_{\hat\vartheta,(\hat m-m)/\sigma,\hat\sigma/\sigma,y}(X,Y)-\varphi_{\vartheta_0,0,1,y}(X,Y))^2\mid \Y_n]=o_P(\delta_n^2)$ uniformly with respect to $y\in\mathbb{R}$ with some $\delta_n\searrow 0$ for $n\to\infty$, where $\Y_n = \{(X_i,Y_i) : i=1,\ldots,n\}$. 
\epro

\noindent
{\bf Proof of Proposition \ref{prop3}}
Note that $\varphi_{\vartheta_0,0,1,y}\equiv 0$.
The expectation in the assertion can be bounded by the sum
\begin{eqnarray}\label{prop3-term1}
&& 2E[(\varphi_{\hat\vartheta,(\hat m-m)/\sigma,\hat\sigma/\sigma,y}(X,Y)-\varphi_{\vartheta_0,(\hat m-m)/\sigma,\hat\sigma/\sigma,y}(X,Y))^2\mid \Y_n]\\
&&{}+2E[(\varphi_{\vartheta_0,(\hat m-m)/\sigma,\hat\sigma/\sigma,y}(X,Y))^2\mid \Y_n].\label{prop3-term2}
\end{eqnarray}
We first consider (\ref{prop3-term1}) which equals
\begin{eqnarray*}
&&E[(I\{\Lambda_{\hat\vartheta}(Y)\leq y\hat\sigma(X)+\hat m(X)\}-I\{\Lambda_{\vartheta_0}(Y)\leq y\hat\sigma(X)+\hat m(X)\})^2\mid \Y_n]\\
&\leq& \int |F_{Y|X}(V_{\hat\vartheta}(y\hat\sigma(x)+\hat m(x))|x)-F_{Y|X}(V_{\vartheta_0}(y\hat\sigma(x)+\hat m(x))|x)|\,dF_X(x)
\end{eqnarray*}
with the notations from the proof of  Theorem \ref{theo1}. Note that this term is very similar to $A_n$ in that proof, only that an absolute value is added inside the integral. With the same methods as there the rate $O_P(n^{-1/2})$ can be shown. 

Next we consider (\ref{prop3-term2}) which equals
\begin{eqnarray*}
&&E\Big[\Big(I\Big\{\eps\leq y\frac{\hat\sigma(X)}{\sigma(X)}+\frac{\hat m(X)-m(X)}{\sigma(X)}\Big\}-I\{\eps\leq y\}\Big)^2\mid \Y_n\Big]\\
&\leq& \int \Big|F_\eps \Big(y\frac{\hat\sigma(x)}{\sigma(x)}+\frac{\hat m(x)-m(x)}{\sigma(x)}\Big)-F_\eps(y)\Big|\,dF_X(x)\\
&\leq& \sup_{y\in\mathbb{R}}|f_\eps(\xi_n(y))|\int \Big|\frac{\hat m(x)-m(x)}{\sigma(x)}\Big|\,dF_X(x)
+\sup_{y\in\mathbb{R}}|yf_\eps(\xi_n(y))|\int \Big|\frac{\hat \sigma(x)-\sigma(x)}{\sigma(x)}\Big|\,dF_X(x)
\end{eqnarray*}
where $\xi_n(y)$ converges to $y$ in probability. Hence the supremum terms are bounded thanks to assumption (A3). Further using the decomposition 
$\hat m-m=(\hat m_{\vartheta_0}-m)+(\hat m_{\hat\vartheta}-\hat m_{\vartheta_0})$ as in the proof of Proposition \ref{prop2} (and similar for $\hat\sigma$) one can show the rate $ O_P((nh^d/\log n)^{-1/2})+ O_P(n^{-1/2})$. This proves the assertion. 
 \hfill $\Box$

\bigskip

\end{appendix}

\bigskip
\bigskip

\begin{center}
{\Large \bf Acknowledgments}
\end{center}
The first author acknowledges financial support by the DFG (Research Unit FOR 1735 {\it Structural Inference in Statistics: Adaptation and Effciency}).  The research of the second author was supported by Basic Science Research Program through the National Research Foundation of Korea (NRF) funded by the Ministry of Education (2014R1A1A2059875). The research of the third author was supported by the European Research Council under the European Community's Seventh Framework Programme (FP7/2007-2013) / ERC Grant agreement No.\ 203650, by IAP research network grant nr. P7/06 of the Belgian government (Belgian Science Policy), and by the contract 'Projet d'Actions de Recherche Concert\'ees' (ARC) 11/16-039 of the 'Communaut\'e fran\c{c}aise de Belgique', granted by the 'Acad\'emie universitaire Louvain'.

\bigskip 

\begin{center}
{\Large \bf References}
\end{center}

\bib Akritas, M. G. and Van Keilegom, I. (2001).
Non-parametric estimation of the residual distribution.
{\it Scand. J. Statist.}  {\bf 28}, 549--567.

\bib Azzalini, A. (2005). The skew-normal Distribution and Related Multivariate Families. {\it Scand. J. Statist.} {\bf 32}, 159--188.

\bib  Bickel, P. J. and Doksum, K. A. (1981).
An analysis of transformations revisited. 
{\it J. Amer. Statist. Assoc.}  {\bf 76}, 296--311.

\bib Birke, M. and Neumeyer, N. (2013).
Testing monotonicity of regression funtions -- an empirical process approach. 
{\it Scand. J. Statist.}  {\bf 40}, 438--454.

\bib Box, G. E. P. and Cox, D. R. (1964).
An analysis of transformations.
{\it J. Roy. Statist. Soc. Ser. B}  {\bf 26}, 211--252.

\bib  Carroll, R. J. and Ruppert, D. (1988).
\textit{Transformation and Weighting in Regression}. 
Monographs on Statistics and Applied Probability. 
Chapman \& Hall, New York.

\bib Colling, B. and Van Keilegom, I. (2014).
Goodness-of-fit tests in semiparametric transformation models. 
Technical report DP2014/17, Universit\'e catholique de Louvain, Institut de Statistique, Biostatistique et Sciences Actuarielles.
\\ (\verb;http://www.uclouvain.be/en-369695.html#DP_2014;) 

\bib Dette, H., von Lieres und Wilkau, C. and Sperlich, S. (2005).
A comparison of different nonparametric methods for inference on additive models.
{\it J. Nonparametr. Stat.}  {\bf 17}, 57--81.

\bib Efromovich, S. (1999). 
{\it Nonparametric Curve Estimation.} Springer, New York. 

\bib Einmahl, J. H. J. and Van Keilegom, I. (2008).
Specification tests in nonparametric regression.
{\it J. Econometrics}  {\bf 143}, 88--102.

\bib Fan, C. and Fine, J. P. (2013).
Linear transformation model with parametric covariate transformations. 
{\it J. Amer. Statist. Assoc.}  {\bf 108}, 701--712.

\bib Fan, J. and Gijbels, I. (1996). 
\textit{Local Polynomial Modelling and Its Applications}. 
Chapman \& Hall, London.

\bib Gijbels, I., Omelka, M. and Veraverbeke, N. (2013).
Estimation of a copula when a covariate affects only marginal distributions.
Technical report. \\
(\verb;http://iap-studys.be/publications/technicalreports/2013;)

\bib Gonz\'{a}lez-Manteiga, W. and Crujeiras, R. M. (2013).
An updated review of goodness-of-fit tests for regression models.
{\it TEST}  {\bf 22}, 361--411.

\bib Gu, J., Li, Q. and Yang, J-C. (2014).
 Multivariate local polynomial kernel estimators: leading bias and asymptotic distribution.
{\it Econometric Reviews}, to appear. 

\bib H\"{a}rdle, W. and Mammen, E. (1993).
Comparing nonparametric versus parametric regression fits. 
{\it Ann. Statist.}  {\bf 21}, 1926--1947.

\bib Heuchenne, C., Samb, R. and Van Keilegom, I. (2014). 
Estimating the residuals distribution in semiparametric transformation models.
Technical report DP2014/11, Universit\'e catholique de Louvain, Institut de Statistique, Biostatistique et Sciences Actuarielles.
\\ (\verb;http://www.uclouvain.be/en-369695.html#DP_2014;) 

\bib Hl\'{a}vka, Z., Hu\v{s}kov\'{a}, M. and Meintanis, S. G. (2011).
Test for independence in non-parametric heteroscedastic regression models.
{\it J. Multivariate Anal.}  {\bf 102}, 816--827.

\bib Horowitz, J. L. (1996).
Semiparametric estimation of a regression model with an unknown transformation of the dependent variable.
{\it Econometrica}  {\bf 64}, 103--137.

\bib Horowitz, J. L. (2009).
\textit{Semiparametric and nonparametric methods in econometrics}.
Springer Series in Statistics. Springer, New York.

\bib Hu\v{s}kov\'{a}, M. and Meintanis, S. G. (2010). 
Tests for the error distribution in nonparametric possibly heteroscedastic regression models.
{\it TEST}  {\bf 19}, 92--112.

\bib Linton, O., Sperlich, S. and Van Keilegom, I. (2008).
Estimation on a semiparametric transformation model.
{\it Ann. Statist.}  {\bf 36}, 686--718.

\bib Masry, E. (1996a).
Multivariate local polynomial regression for time series: uniform strong consistency and rates. 
{\it J. Time Ser. Anal.}  {\bf 17}, 571--599. 

\bib Masry, E. (1996b).
Multivariate regression estimation -- local polynomial fitting for time series. 
{\it Stochastic Process. Appl.}  {\bf 65}, 81--101.

\bib Mu, Y. and He, X. (2007).
Power transformation toward a linear regression quantile.
{\it J. Amer. Statist. Assoc.}  {\bf 102}, 269--279.

\bib Neumeyer, N. (2009a). 
Testing independence in nonparametric regression.
{\it J. Multivariate Anal.} {\bf 100},  1551--1566.

\bib Neumeyer, N. (2009b). 
Smooth residual bootstrap for empirical processes of nonparametric regression residuals. 
{\it Scand. J. Statist.}  {\bf 36}, 204--228.

\bib Neumeyer, N. and Sperlich, S. (2006).
Comparison of separable components in different samples.
{\it Scand. J. Statist.}  {\bf 33}, 477--501.

\bib Neumeyer, N. and Van Keilegom, I. (2009). 
Change-point tests for the error distribution in nonparametric regression. 
{\it Scand. J. Statist.}  {\bf 36}, 518--541.

\bib Neumeyer, N. and Van Keilegom, I. (2010). 
Estimating the error distribution in nonparametric multiple regression with applications to model testing. 
{\it J. Multivariate Anal.}  {\bf 101}, 1067--1078.

\bib Ruppert, D., Sheather, S. J. and Wand, M. P. (1995). An effective bandwidth selector for local least
squares regression. {\it J. Amer. Statist. Assoc.} {\bf 90}, 1257--1270.

\bib Sheather, S. J. and Jones, M. C. (1991). A reliable data-based bandwidth selection method for
kernel density estimation. {\it J. Roy. Statist. Soc. Ser. B} {\bf 53}, 683--690.

\bib Stute, W., Gonz\'{a}lez-Manteiga, W. and Presedo Quindimil, M. (1998).
Bootstrap approximations in model checks for regression. 
{\it J. Amer. Statist. Assoc.}  {\bf 93}, 141--149.

\bib Van der Vaart, A. W. (1998).  
\textit{Asymptotic Statistics}.  
Cambridge University Press. Cambridge.

\bib Van der Vaart, A. W. and Wellner, J. A. (1996).
\textit{Weak Convergence and Empirical Processes}.
Springer-Verlag, New York.


\bib Yeo, I-K. and Johnson, R. A. (2000).
A new family of power transformations to improve normality or symmetry. 
{\it Biometrika}  {\bf 87}, 954--959.

\bib Zhou, X-H., Lin, H. and Johnson, E. (2008).
Non-parametric heteroscedastic transformation regression models for skewed data with an application to health care costs.
{\it J. Roy. Statist. Soc. Ser. B}  {\bf 70}, 1029--1047. 

\bib Zhu, L., Fujikoshi, Y. and Naito, K. (2001).
Heteroscedasticity checks for regression models. 
{\it Sci. China Ser. A} {\bf 44}, 1236--1252.

\end{document}